\documentclass[preprint,12pt]{elsarticle}

\usepackage[T1]{fontenc}
\usepackage[utf8]{inputenc}
\usepackage{amssymb}
\usepackage{amsmath}
\usepackage{cancel} 
\usepackage{subfig}
\usepackage{pgf}  
\usepackage{multirow}
\usepackage{todonotes}
\usepackage{hyperref}
\usepackage{hyphenat}
\usepackage{siunitx}
\usepackage{float}
\usepackage{algorithm}
\usepackage{algpseudocode}
\usepackage{textpos}
\usepackage{mdframed}
\usepackage{enumitem}
\usepackage{fvextra}
\usepackage{inconsolata} 
\usepackage{afterpage} 
\usepackage{makecell}
\usepackage{siunitx}
\usetikzlibrary{shapes.geometric, arrows.meta, plotmarks, positioning}

\newlength{\extralength}
\usepackage{booktabs}
\usepackage{changepage}
\usepackage{multicol}

\usepackage[a4paper, left=2cm, right=2cm, top=1.5cm, bottom=3cm]{geometry}

\graphicspath{{./figs/}}

\newcommand{\drawHistogram}[6]{

    \pgfmathsetmacro{\sumValues}{#1 + #2 + #3 + #4 + #5 + #6}

    \pgfmathsetmacro{\scaleFactor}{2}

    \pgfmathsetmacro{\normA}{(#1 / \sumValues) * \scaleFactor}
    \pgfmathsetmacro{\normB}{(#2 / \sumValues) * \scaleFactor}
    \pgfmathsetmacro{\normC}{(#3 / \sumValues) * \scaleFactor}
    \pgfmathsetmacro{\normD}{(#4 / \sumValues) * \scaleFactor}
    \pgfmathsetmacro{\normE}{(#5 / \sumValues) * \scaleFactor}
    \pgfmathsetmacro{\normF}{(#6 / \sumValues) * \scaleFactor}

    \begin{tikzpicture}
        \path (-0.25,-0.25) (5.25,1.75 * \scaleFactor);
        \draw[gray,line width=4.5pt] plot coordinates {
            (0,0)
            (0, \normA)
            (1, \normA)
            (1, \normB)
            (2, \normB)
            (2, \normC)
            (3, \normC)
            (3, \normD)
            (4, \normD)
            (4, \normE)
            (5, \normE)
            (5, \normF)
            (6, \normF)
            (6, 0)
        };
        \foreach \i in {0,...,6} {
            \path plot[mark=square*,mark options={color=black},mark size=1.5pt] coordinates { (\i, 0) };
        }

    \end{tikzpicture}
}

\makeatletter
\def\ps@pprintTitle{   \let\@oddhead\@empty
   \let\@evenhead\@empty
   \def\@oddfoot{\reset@font\hfil\thepage\hfil}
   \let\@evenfoot\@oddfoot
}
\makeatother

\graphicspath{{./figs/}}

\begin{document}

\begin{frontmatter}

\title{DeepSeek-V3, GPT-4, Phi-4, and LLaMA-3.3 generate correct code for LoRaWAN-related engineering tasks}

\author[label1]{Daniel Fernandes}
\author[label2,label3]{João P. Matos-Carvalho}
\author[label1,label3]{Carlos M. Fernandes}
\author[label1,label3]{Nuno Fachada}

\affiliation[label1]{organization={Lusófona University, COPELABS},
            city={Lisboa},
            postcode={1749-024},
            country={Portugal}}

\affiliation[label2]{organization={LASIGE, Departamento de Informática, Faculdade de Ciências, Universidade de Lisboa, 1749--016 Lisboa, Portugal},
}

\affiliation[label3]{organization={Center of Technology and Systems (UNINOVA-CTS) and Associated Lab of Intelligent Systems (LASI)},
            city={Caparica},
            postcode={2829-516},
            country={Portugal}}

\begin{abstract}
    This paper investigates the performance of 16 Large Language Models (LLMs) in automating LoRaWAN-related engineering tasks involving optimal placement of drones and received power calculation under progressively complex zero-shot, natural language prompts. The primary research question is whether lightweight, locally executed LLMs can generate correct Python code for these tasks. To assess this, we compared locally run models against state-of-the-art alternatives, such as GPT-4 and DeepSeek-V3, which served as reference points. By extracting and executing the Python functions generated by each model, we evaluated their outputs on a zero-to-five scale. Results show that while DeepSeek-V3 and GPT-4 consistently provided accurate solutions, certain smaller models---particularly Phi-4 and LLaMA-3.3---also demonstrated strong performance, underscoring the viability of lightweight alternatives. Other models exhibited errors stemming from incomplete understanding or syntactic issues. These findings illustrate the potential of LLM-based approaches for specialized engineering applications while highlighting the need for careful model selection, rigorous prompt design, and targeted domain fine-tuning to achieve reliable outcomes.
\end{abstract}

\begin{keyword}
LoRaWAN \sep Large Language Models \sep UAV Placement \sep Code Generation \sep IoT

\end{keyword}

\end{frontmatter}

\begin{textblock*}{190mm}(-1cm,-17.18cm)
    {\noindent \footnotesize \color{black!60} The peer-reviewed version of this paper is
    published in Electronics at \url{https://doi.org/10.3390/electronics14071428}.
    This version is typeset by the authors and differs only in pagination and
    typographical detail.}
\end{textblock*}

\section{Introduction}

The rapid expansion of Internet of Things (IoT) applications has led to increased attention to Low-Power Wide-Area Network (LPWAN) technologies, such as LoRa Wide Area Network (LoRaWAN), which provide long-range communication with low power consumption~\cite{petajajarvi2015coverage}. LoRaWAN networks are particularly appealing for rural areas, where infrastructure constraints can pose significant challenges to traditional wireless communication systems~\cite{augustin2016study}. In~this context, the~integration of Unmanned Aerial Vehicles (UAVs) as mobile relays has emerged as a promising solution, enabling flexible deployments and extended coverage~\cite{mozaffari2017wireless}. Determining the UAV position that minimizes signal propagation loss and assessing the corresponding received power are critical for ensuring reliable connectivity and resource-efficient operations in these rural scenarios~\cite{sanchez2018performance}.

Parallel to these developments in wireless communications, Large Language Models (LLMs) have shown rapid progress. Modern LLMs---including GPT-4~\cite{achiam2024gpt}, recent open-source offerings locally installable with Ollama~\cite{ollama2025}, and~novel models such as DeepSeek~\cite{liu2024deepseekv3}---have shown substantial capabilities in understanding complex tasks and generating functional code for engineering problems~\cite{achiam2024gpt}. Furthermore, these models demonstrate a broad applicability beyond code generation, including text clustering~\cite{petukhova2025text}, text summarization~\cite{lewis2019bart}, machine translation~\cite{alves2024tower}, and~text classification/question answering~\cite{yang2019xlnet}. However, despite these advancements, the~effectiveness of lightweight, locally executed models in generating correct and efficient solutions for domain-specific engineering tasks remains an open question~\cite{gu2024effectiveness}.

This study investigates whether lightweight and locally executed LLMs can generate correct Python code for UAV planning tasks in LoRaWAN environments. Specifically, we assess 16 different LLMs by evaluating their ability to generate Python functions that determine the optimal UAV position from a discrete set of candidate locations, minimizing propagation loss, and~computing the corresponding received power (in dBm). Our primary goal is to compare the performance of locally run models, such as LLaMA-3.3~\cite{grattafiori2024llama} and Phi-4~\cite{abdin2024phi}, against~state-of-the-art large models such as GPT-4~\cite{achiam2024gpt} and DeepSeek-V3~\cite{liu2024deepseekv3}, accessed via their online application programming interfaces (APIs). The~inclusion of these larger models serves as a reference point to establish that such tasks can indeed be solved using advanced LLMs, allowing for a meaningful comparison with the performance of smaller, locally executed alternatives. The~evaluation uses a zero-shot natural language prompt configuration, and~correctness is measured through a scoring system based on function extraction and execution~results.

Despite significant progress in AI-assisted UAV deployment, previous research has largely overlooked the unique communication and operational constraints inherent to LoRaWAN environments. LoRaWAN deployments pose distinct challenges such as stringent power limitations, specialized propagation characteristics at lower frequencies, and~long-range communication requirements that differ fundamentally from scenarios commonly studied in existing UAV-AI literature. Existing approaches primarily focus on UAV trajectory planning, mission coordination, or~visual scene understanding tasks, without~explicitly addressing scenarios involving the low-power, wide-area network constraints and signal propagation peculiarities of LoRaWAN systems. This gap motivates our study, which specifically examines whether LLMs---particularly lightweight, locally executable variants---can effectively generate Python code to solve UAV placement and received power calculation tasks uniquely relevant to LoRaWAN~environments.

The findings of this study are significant for two main reasons. First, they illustrate the extent to which lightweight, locally run LLMs can perform domain-specific engineering tasks, providing insight into their potential as cost-effective alternatives to proprietary, large-scale models~\cite{ling2024domain}.
Second, these findings may offer practical guidance not only for practitioners integrating LLM-generated code into IoT and UAV communication workflows but also for those in a wide range of other fields, as~they highlight critical considerations such as reliability, correctness, and~maintainability.
The subsequent sections of this paper are organized as follows. Section~\ref{sec:background} provides background information on the use of LLMs for human--UAV interaction and code generation, also discussing relevant aspects of prompt design. Section~\ref{sec:methods} describes the materials and methods employed, including the engineering problem context, prompt structure, model selection, and~evaluation metrics. Results are presented in Section~\ref{sec:results}, followed by a detailed discussion in Section~\ref{sec:discussion}. Section~\ref{sec:limitations} outlines the study's limitations and opportunities for future research. Finally, Section~\ref{sec:conclusions} concludes the paper with final remarks and~recommendations.

\section{Background}
\label{sec:background}

In this section, we start by addressing the general goal of integrating LLMs with UAVs to improve the behavior, organization, and~communication of autonomous systems, as~well as the specific implementation of UAVs as mobile relays and antennas in LoRoWAN environments. In~Section~\ref{sec:llm_code}, we focus on the specific task of generating code for autonomous devices and on how LLMs are being used to incorporate code generation at different levels of workflow. Finally, in~Section~\ref{sec:prompt_design}, we briefly discuss prompt engineering and its principles, the~benefits and drawbacks of conversational and structured prompting, and~how prompt design impacts code generation or task planning.
\
\subsection{LLMs for Human--UAV~Interaction}
\label{sec:llm_human_drone}

The nature of UAVs, namely their collective organization and communication requirements, strongly encourages integration with Artificial Intelligence (AI) algorithms. The~recent emergence of LLM technologies in particular is inspiring new frameworks and prototypes for communication and design of several autonomous systems, and UAVs are no exception. As~LLMs learning and adaptation capabilities in uncertain and dynamic environments grow and approach human-level proficiency, the~scientific literature on the subject steadily increases~\cite{javaid2024large,phadke2024integrating}. Currently, there is a significant amount of knowledge on LLMs for human--UAV interaction. For~a review on the state-of-the art literature on LLMs and UAVs, please refer to~\cite{javaid2024large}. For~a discussion of key areas where LLMs can impact UAVs, we urge the reader to refer to the paper by Phadke~et~al.~\cite{phadke2024integrating}. In~the following paragraphs, we discuss some recent developments on the usage of natural language models for controlling~UAVs.

In~\cite{electronics13224508}, Aikins~et~al. present LEVIOSA, a~framework for the generation of UAV trajectory based on text and speech. The~authors use several LLMs to convert natural language prompts into sets of coordinates to guide the UAVs and low-level controllers to control each device in its path, aiming for accuracy, synchronization, and~collision avoidance.  LEVIOSA was tested on various scenarios with promising~results.

Cui~et~al.~\cite{cui2024tpml} propose a Task Planning for Multi-UAV System (TPML) that uses LLMs as interfaces to translate UAV’s operator instructions into executable codes. After~validating the system in simulation environments and real-wold scenarios, the~authors argue that TPML is able to control multiple UAVs in both synchronous and asynchronous missions with a single natural language~input.

While most of the studies on natural language processing for UAVs focus on processing the user messages to program or optimize UAV behavior, others try to provide UVAs with scene descriptions skills in natural language, taking advantage of their capacity to acquire visual cues of the environment. In~\cite{drones7020114}, the~authors use LLMs and Visual Language Models (VLMs) to provide UAVs with the ability of scene description using natural language. The~generated tests were subject to a readability test, some achieving a high school senior reading level (level 12 in the Gunning fog index).

In~\cite{zhu2024task}, the~authors discuss a framework that integrates a novel factorization method---QTRAN---in a multi-agent reinforcement learning algorithm (MARL)~\cite{son2019qtran} with an LLM to optimize UAV trajectories, overcoming limitations of value decomposition algorithms for trajectory planning, as~they have difficulties in associating local observations with the global state of UAV swarms. Although~QTRAN overcomes some of the limitations of standard MARLs, its performance can still be improved, namely by enhancing the representation network. For~that purpose, the~authors incorporate LLMs in the framework, boosting its overall performance in trajectory optimization and outperforming other reinforcement learning~methods.

LPWAN-based systems are one of the emerging technologies in which UAVs are being tested and deployed. LPWANS, and~LoROWANs in particular, rely on a set of fixed sensor stations, which measure and transmit a number of environmental data to a central unit. Traditionally, these stations are static, cover only very small areas and can be impaired by natural disasters. Due to their mobility, UAVs can act as moving communication nodes, which solves some of the limitations of static~LoROWANs.

Several methods have been proposed to integrate UAVs in LoROWANs. In~\cite{s20082396}, UAVs are used to transfer information from ground-based LORAWAN nodes to the base station. The~architecture of the systems thus consists of two layers, the~first being the ground nodes that transmit data using LoRaWAN and the second the swarm of drones communicating over a WiFi ad~hoc network. To~enhance the performance of the systems, a~distributed topology algorithm periodically adapts the UAV topology to the position of the ground nodes. In~\cite{8519967}, the~authors describe an air quality monitor system based on a LORAWAN and UAVs. In~\cite{9477431}, a~UAV emergency monitoring system using a LORAWAN is proposed to overcome the limitations of ground stations in disaster scenarios. Finally, Arroyo~et~al.~\cite{electronics11121865} propose a UAV and LOROWAN system that enables data transfer from sensors to a central system and then use machine learning to classify the data. To~the extent of our knowledge, there are no studies on the integration of LLMs and UAVs in a LoRaWAN~environment.

\subsection{Code Generation with~LLMs}
\label{sec:llm_code}

The landscape of AI-assisted programming has evolved significantly, with~extensive research focusing on natural language generation and understanding of large codebases~\cite{wong2023natural}. Shortly after their inception, some LLMs demonstrated capabilities in code assistance and code generation, even from natural language specifications. In~the first models, those skills were somewhat limited and the output often required post-processing steps to improve the quality of the suggested code~\cite{10.1145/3510003.3510203}. But~LLMs quickly evolved, and~their ability to provide executable code in due time improved significantly~\cite{brown2020language}. Furthermore, derivations of popular LLMs, like Open AI Codex~\cite{chen2021evaluating}, a~descendant of ChatGPT-3, and~Code Llama~\cite{grattafiori2024llama}, Meta's programming tool, emerged as specialized models for coding. Nowadays, AI-assisted programming is a common practice in~industry.

In the context of code generation for autonomous devices, Vemprala~et~al.~\cite{vemprala2023chatgpt} explore ChatGPT's ability on several robot-oriented tasks, including code synthesis. The~authors present a framework for robot control that requires designing and implementing a library of APIs receptive to prompt engineering for ChatGPT. The~proposed framework allows the generated code to be tested, verified, and~validated by a user through simulation and manual~inspection.

In~\cite{10160591}, the~authors adapt LLMs trained on code completion for writing robot policy code according to natural language prompts. The~generated robot policies exhibit spatial-geometric reasoning and are able to prescribe precise values to ambiguous descriptions. By~relying on a hierarchical prompting strategy, their approach is able to write more complex code and solve 39.8\% of the problems on the HumanEval~\cite{chen2021evaluating} benchmark.

Luo~et~al.~\cite{LUO2024100488} use LLMs to generate robot control programs, testing and optimizing the output in a simulation environment. After~a number of optimization rounds, the~robot control codes are deployed on a real robot for construction assembly tasks. The~experiments show that their approach can improve the quality of the generated code, thus simplifying the robot control process and facilitating the automation of construction~tasks.

\subsection{Prompt~Design}
\label{sec:prompt_design}

The piece of text or set of instructions that the user provides to an LLM to generate a specific response is called a prompt. Designing effective prompts is essential to take advantage of the potential of LLMs, and~in a few years the craft established as a field of research and development of its own~\cite{amatriain2024prompt}.

Prompting strategies can be broadly classified into structured and unstructured approaches. Structured prompting employs precise instructions with explicitly defined inputs, outputs, and~constraints, often leading to more reliable and accurate code generation. However, structured prompts typically require a deeper understanding of both the problem domain and the underlying model, potentially limiting flexibility and accessibility. Conversely, unstructured prompting uses intuitive, conversational language, making it accessible to a broader audience, reflecting realistic scenarios where users may not possess specialized knowledge of prompt crafting. However, this can result in less consistent outputs due to inherent~ambiguity.

Prompts may also be categorized based on the number of illustrative examples provided: zero-shot prompts provide no examples, one-shot prompts include a single example, and~few-shot prompts incorporate multiple examples. Empirical research supports the trade-offs associated with different prompt styles; for instance, Liang~et~al.~\cite{10160591} demonstrate that structured, code-based prompts generally yield superior results for robot-related reasoning tasks compared to natural language prompts. However, advances in LLM technology continue to improve the viability of unstructured, natural language prompting in complex domains such as robotics~\cite{vemprala2023chatgpt}. Further improvements in output coherence have also been observed through structured reasoning techniques such as chain-of-thought (CoT) prompting~\cite{10.5555/3600270.3602070, LUO2024100488}.

In this study, we follow a natural language zero-shot prompt strategy, in~which the request is performed in a relatively unstructured fashion without any examples. Nonetheless, established best practices for engineering-focused code generation were followed by explicitly specifying function inputs, expected return types, and~required libraries, thus improving the clarity and reproducibility of the generated code~\cite{li2024approach}.

\section{Materials and~Methods}
\label{sec:methods}

This section starts with an overview of the theoretical context that informs our prompt design in Section~\ref{sec:methods:theory}. Next, Section~\ref{sec:methods:scenarios} presents the proposed prompts and their respective scenarios. Section~\ref{sec:methods:llms} describes and justifies the models analyzed in this study. Section~\ref{sec:methods:impl} then outlines the prompting and response processing pipeline. The~section concludes with a description of the experimental setup in Section~\ref{sec:methods:expsetup}, including all tested inputs for both the LLMs and the generated Python functions, the~expected function results, and~the evaluation metrics~used.

\subsection{Theoretical~Context}
\label{sec:methods:theory}

The IoT paradigm refers to the interconnection of physical devices that collect, exchange, and~process data over the Internet or other communication networks. According to Sanguesa~et~al.~\cite{sanguesa_improving_2023}, it is estimated that by 2030, there will be approximately 125 billion IoT devices, ranging from simple temperature and humidity sensors to more complex sensors used in sectors such as agriculture and industry. The~main goal of these sensors is to simplify and optimize daily activities. One of the challenges associated with this paradigm is the large volume of data generated and how it is processed. A~potential solution for data collection is the use of UAVs, which can fly over (or carry) multiple sensors along a predefined path planning. These UAVs may or may not be capable of transmitting data in real time to a base station (BS). However, to~use UAVs efficiently, it is often necessary to calculate their location and send control commands to adjust their position or even modify their flight path. Therefore, reliable communication between the UAV and a base station is crucial. One possible communication protocol for this purpose is LoRaWAN, which is based on LoRa (long-range) communication and enables effective long-distance data transmission~\cite{raimundo_uav_nodate,ghazali_systematic_2021}. Essentially, LoRa communication establishes a link between two points: the transmitter---in this case, the~BS---and the receiver, i.e.,~the UAV. This communication is based on classical propagation models, such as those found in reference~\cite{saunders_antennas_2007}.

Regarding the modulation of a communication channel, the~received power at the antenna ($p_r$) depends on factors such as the transmit power ($p_t$), the~gain of the antennas ($g_r$ and $g_t$), the~distance between the antennas ($r$), and~the losses during transmission (free-space attenuation). Equation~(\ref{equ:freespace}) represents the propagation loss $l_F$ between the \mbox{two points:}

\begin{equation}
\label{equ:freespace}
    l_F = \frac{p_t \cdot g_r \cdot g_t }{p_r} =
    \left( \frac{4 \pi r}{\lambda} \right)^2 =
    \left( \frac{4 \pi r f }{c} \right)^2
\end{equation}

\noindent where $\lambda$ represents the wavelength. In~particular, $\lambda = \frac{c}{f}$, with~$c$ representing the speed of light and $f$ the frequency, which in Europe is $868$ MHz.

A lower propagation loss results in a stronger received signal. Propagation losses are typically expressed in $dB$ units, and~for a distance in meters and a frequency in $Hz$, Equation~(\ref{equ:freespace}) can be rewritten as Equation~(\ref{equ:freespace2}), which represents the Free Space Path Loss formula. This formula is valid under free-space conditions, assuming a direct, unobstructed line of sight. In~terms of notation, lowercase variables denote linear values, whereas uppercase variables denote logarithmic values.
\begin{equation}
\label{equ:freespace2}
    L_F{(dB)} = 20\log(r_{m}) + 20\log(f_{Hz}) - 147.55
\end{equation}

To estimate the received power, it is necessary to consider the transmitted power, the~gain of the transmitting and receiving antennas, and~the path losses that occur during transmission. Thus, Equation~(\ref{equ:rx}), derived from Equation~(\ref{equ:freespace}), can be written as
\begin{equation}
\label{equ:rx}
    P_{r(dBm)} = P_t + G_t + G_r - L_F
\end{equation}

\subsection{Scenarios and~Prompts}
\label{sec:methods:scenarios}

To evaluate the LLM models, three zero-shot prompts with increasing levels of difficulty were designed---see Table~\ref{tab:prompts}. In~this context, `zero-shot' refers to prompts that do not provide any examples to the model being tested. Furthermore, these prompts use natural language, meaning that they are relatively unstructured and have undergone minimal refinement, apart from ensuring technical precision and clarity. This approach was chosen as it more closely follows real-world scenarios where domain experts may rely on direct, straightforward queries to achieve their~goals.

The specific request posed by these prompts is for the LLM to identify, from~a set of points, the~point where the value of $L_F$ is the lowest or to determine the received power at that point (i.e., the~point with the lowest $L_F$). In~all scenarios, a~frequency of $868$ MHz is considered, as~well as a rural area where LoRa communication is possible up to $10$ km. Both antennas are assumed to have a gain of $2.5$ dBi~each.

To simplify post-processing of responses, all prompts specify the available libraries, the~expected indentation type, and~that the return function should be self-contained---i.e., all required code including constants and auxiliary functions should be defined within the requested~function.

\begin{table}
    \caption{Prompts designed for this study, requiring the tested LLMs to generate Python functions that solve increasingly complex tasks related to LoRaWAN and UAVs.}
    \label{tab:prompts}
    \centering
    \begin{tabular}{l}

        \toprule
        \footnotesize{\textbf{Prompt 1}}\\
        \midrule

\begin{minipage}{17cm}
\begin{Verbatim}[fontsize=\scriptsize,breaklines,breaksymbolleft={}, breaksymbolright={}, breakindent=0pt]
Consider that the LoRaWAN communication protocol is being used in a rural scenario where a base station communicates with a UAV at a communication frequency of 868 MHz. Assume a system with two axes (the x-axis and the y-axis) and that the base station is in position (0,0). Also, assume that all positions are in kilometers (km).

Create a Python function called `index_position()` which accepts a list of tuples, with each (x, y) tuple representing a possible position in which the UAV can be placed with respect to the base station. This function should return the list index of the tuple (i.e. UAV position) which minimizes the propagation loss. Assume that the math and numpy libraries are imported as follows, and no more libraries can be used:

import math
import numpy as np

Beyond importing these libraries, the `index_position()` function must be self-contained. In other words, all variables, constants, or helper functions must be defined within the `index_position()` function. Provide Python code with 4-space indentation following PEP 8.
\end{Verbatim}
\end{minipage} \\

        \midrule
        \footnotesize{\textbf{Prompt 2}}\\
        \midrule

\begin{minipage}{17cm}
\begin{Verbatim}[fontsize=\scriptsize,breaklines,breaksymbolleft={}, breaksymbolright={}, breakindent=0pt]
Consider that the LoRaWAN communication protocol is being used in a rural scenario where a base station communicates with a UAV at a communication frequency of 868 MHz. Assume a system with two axes (the latitude axis and the longitude axis) where each value is given in decimal degrees.

Create a Python function called `index_position()` which accepts a list of (latitude, longitude) tuples. The first tuple in this list represents the position of the base station, while the remaining tuples represent possible positions in which the UAV can be placed. This function should return the list index of the tuple which minimizes the propagation loss. Assume that the math and numpy libraries are imported as follows, and no more libraries can be used:

import math
import numpy as np

Beyond importing these libraries, the `index_position()` function must be self-contained. In other words, all variables, constants, or helper functions must be defined within the `index_position()` function. Provide Python code with 4-space indentation following PEP 8.
\end{Verbatim}
\end{minipage} \\

        \midrule
        \footnotesize{\textbf{Prompt 3}}\\
        \midrule

\begin{minipage}{17cm}
\begin{Verbatim}[fontsize=\scriptsize,breaklines,breaksymbolleft={}, breaksymbolright={}, breakindent=0pt]
Consider that the LoRaWAN communication protocol is being used in a rural scenario where a base station communicates with a UAV at a communication frequency of 868 MHz, with a transmission power of 27 dBm. Both the transmitter and UAV antennas have a gain of 2.5 dBi. Assume a system with two axes (the latitude axis and the longitude axis) where each value is given in decimal degrees.

Create a Python function called `power_received()` which accepts a list of (latitude, longitude) tuples. The first tuple in this list represents the position of the base station, while the remaining tuples represent possible positions in which the UAV can be placed. This function should return the power received (in dBm) by the UAV at the position that minimizes the propagation loss. Assume that the math and numpy libraries are imported as follows, and no more libraries can be used:

import math
import numpy as np

Beyond importing these libraries, the `power_received()` function must be self-contained. In other words, all variables, constants, or helper functions must be defined within the `power_received()` function. Provide Python code with 4-space indentation following PEP 8.

\end{Verbatim}
\end{minipage} \\

    \bottomrule
    \end{tabular}
\end{table}

The first prompt is presented in the first row of Table~\ref{tab:prompts}. In~this simpler scenario, the~BS and the UAV's possible positions, measured in kilometers (km), are defined within a coordinate system with two axes: the $x$-axis and the $y$-axis. The~BS is fixed at position $(0,0)$, while the UAV's possible positions are provided as an input array to the function generated by the LLMs. To~solve this problem, LLMs must generate a Python function that calculates the distance (e.g., Euclidean) between the BS and each possible UAV position, applies Equation~(\ref{equ:freespace2}) to compute power losses, and~returns the index of the position with the lowest loss. The~LLM must ensure that power losses maintain a one-to-one correspondence with the UAV positions to return the correct~index.

Prompt~2, shown in the second row of Table~\ref{tab:prompts}, increases the complexity by considering geographical coordinates—latitude and longitude—instead of a simple $(x,y)$ axis. LLMs must use a different method to calculate the distances between the UAV's position and the BS, such as Haversine's formula. This prompt further increases the difficulty by requiring that the UAV's position be given as the first element of the input array. Consequently, the~generated functions must extract this information and return an index greater than zero, as~index zero contains the UAV's~position.

Prompt~3, presented in the last row of Table~\ref{tab:prompts}, closely resembles Prompt~2. However, instead of returning the index with the lowest loss, the~generated function must return the value of that loss by applying Equation~(\ref{equ:rx}).

\subsection{LLMs~Considered}
\label{sec:methods:llms}

The LLMs models used in this paper were chosen based on their impact in AI research, innovative approaches, and~performance across different domains such as programming, advanced reasoning, and~computational efficiency. Table~\ref{table:llms} lists and characterizes the LLMs selected for this study. For~the remainder of this paper, the~number of parameters associated with each model is expressed in billions or trillions with an uppercase B and T, respectively.

\begin{table}
    \footnotesize
    \caption{Characteristics and main purpose of the LLMs tested in this study. `Size` indicates the number of parameters in billions (B) or trillions (T). `Tag` corresponds to the specific model version invoked in the respective API calls.}
    \label{table:llms}
    \centering

    \begin{tabular}{llrll}
        \toprule
        Family & Version & \multicolumn{1}{l}{Size} & Tag & Main purpose \\
        \midrule
        DeepSeek~\cite{liu2024deepseekv3,guo2025deepseekr1}

        & R1 & 7B & \texttt{deepseek-r1:7b} & \makecell[tl]{Computationally efficient distilled\\ reasoning model.}  \\

        & R1 & 70B & \texttt{deepseek-r1:70b} & \makecell[tl]{Distilled reasoning model balancing performance\\ and computational efficiency.}  \\

        & V3 & 671B & \texttt{deepseek-v3} & \makecell[tl]{Mixture-of-Experts general-purpose model.}  \\

        \midrule
        Gemma~\cite{mesnard2024gemma,riviere2024gemma}

        & 1.1 & 2B & \texttt{gemma:2b} & \makecell[tl]{Lightweight model for dialogue,\\ instruction-following, and coding.}  \\

        & 2.0 & 2B & \texttt{gemma2:2b} & \makecell[tl]{Compact general-purpose model trained\\ with knowledge distillation.}  \\

        \midrule
        GPT~\cite{achiam2024gpt}

        & 4 & 1.76T\textsuperscript{*} & \texttt{gpt-4-0613} & \makecell[tl]{Multimodal model optimized by OpenAI\\ for text, audio, and image processing.}  \\

        \midrule
        LLaMA~\cite{grattafiori2024llama}

        & 3.2 & 3B & \texttt{llama3.2:3b} & \makecell[tl]{Lightweight text-only model for multilingual\\ dialogue and text summarization.}  \\

        & 3.3 & 70B & \texttt{llama3.3:70b} & \makecell[tl]{Text-only model for deeper comprehension\\  multilingual conversation.}  \\

        & code & 7B & \texttt{codellama:7b} & Code-generation model.\\

        \midrule
        Mistral~\cite{jiang2023mistral}

        & 0.3 & 7B & \texttt{mistral:7b} & \makecell[tl]{Efficient model for text and code generation,\\ supports function calling.}  \\

        \midrule
        Phi~\cite{abdin2024phi}

        & 4.0 & 14B & \texttt{phi4:14b} & \makecell[tl]{Reasoning model trained using high-quality\\ synthetic data.}  \\

        \midrule
        Qwen~\cite{yang2025qwen25,hui2025qwen25coder}

        & 2.5-coder & 0.5B & \texttt{qwen2.5-coder:0.5b} & \makecell[tl]{Code generation model.}  \\

        & 2.5-coder & 1.5B & \texttt{qwen2.5-coder:1.5b} & \makecell[tl]{Code generation model.}  \\

        & 2.5-coder & 3B & \texttt{qwen2.5-coder:3b} & \makecell[tl]{Code generation model.}  \\

        & 2.5 & 0.5B & \texttt{qwen2.5:0.5b} & \makecell[tl]{General-purpose language model.}  \\

        & qwq & 32B & \texttt{qwq:32b} & \makecell[tl]{Advanced reasoning model for complex\\problem-solving tasks.}  \\

        \bottomrule
        \multicolumn{5}{l}{\footnotesize{\textsuperscript{*} Unofficial estimate.}}
    \end{tabular}
\end{table}

The DeepSeek family of models includes a range of architectures designed to balance performance and computational efficiency. DeepSeek-R1 (7B) and DeepSeek-R1 (70B) are distilled versions derived from the larger DeepSeek-R1 model (671B)---based on the Qwen and LLaMA architectures---to retain significant reasoning capabilities while reducing hardware demands~\cite{guo2025deepseekr1}. In~contrast, DeepSeek-V3 (671B) is a Mixture-of-Experts model  designed to perform well in diverse tasks~\cite{liu2024deepseekv3}. Considering these models is crucial due to their varied architectures and training methodologies, which offer insights into the trade-offs between model size, training techniques, and~task-specific performance. The~V3 671B model was selected over its more developed R1 counterpart, as~initial trials demonstrated it was sufficiently accurate for the prompts presented in Section~\ref{sec:methods:scenarios}, providing a balance between performance and~cost.

The Gemma model family~\cite{mesnard2024gemma,riviere2024gemma}, developed by Google DeepMind, comprises open models derived from the research and technology behind the Gemini models. While influenced by Gemini, Gemma is fully open-source and designed for efficient language understanding and reasoning. The~lightweight Gemma v1.1 (2B) and Gemma2 (2B) implementations are optimized for resource-limited environments. Gemma2 (2B) incorporates knowledge distillation, improving efficiency and performance relative to its size. These models were included to assess the trade-offs in model scaling, particularly for the real-time and cost-sensitive applications associated with the tested~prompts.

OpenAI's Generative Pre-trained Transformer (GPT) models are proprietary LLMs designed to understand and generate human-like text, facilitating tasks such as drafting documents, coding, and~responding to queries~\cite{achiam2024gpt}. Their popularity and advanced capabilities make them essential subjects in LLM comparison studies. In~this context, GPT-4-0613 was selected over newer models such as GPT-4o and o1, as~preliminary tests indicated its performance was sufficient for the presented prompts, therefore reducing~costs.

The LLaMA series by Meta AI includes models optimized for various applications~\cite{grattafiori2024llama}. LLaMA-3.2 (3B) is a lightweight, multilingual model suited for mobile and edge devices, appropriate for text summarization and classification. LLaMA-3.3 (70B) is a larger, instruction-tuned model with superior performance in natural conversation and multilingual tasks. Code Llama (7B) specializes in code generation and understanding. Testing these three models is important for evaluating how model size, specialization, and~efficiency in the LLaMA family impacts performance across the three implemented~prompts.

The Mistral family of language models~\cite{jiang2023mistral}, developed by the French company Mistral AI, stands out for its efficient architecture and strong performance. Mistral models achieve high accuracy with fewer parameters, making them more accessible and computationally efficient compared to many large-scale models. The~Mistral v0.3 (7B) model exemplifies this approach, demonstrating capabilities in text and code generation, conversation, and~function calling, while effectively handling longer sequences. Its open-source nature offers a valuable option for research and application development, providing a European alternative to models predominantly from U.S.- and China-based~companies.

The Phi model family~\cite{abdin2024phi}, developed by Microsoft Research, is focused on the role of high-quality synthetic data for improving reasoning in compact language models. Phi-4, a~14-billion parameter model, prioritizes synthetic data to improve problem-solving in mathematics and coding, outperforming its teacher model, GPT-4, on~several benchmarks. Unlike models that primarily scale with size, Phi-4 follows a distinct training approach, making it important to compare against other LLMs. Its relatively small size also makes it relevant for low-resource environments, where optimizing data efficiency can be a crucial factor in model~deployment.

The Qwen model family, developed by Alibaba Cloud, includes general-purpose~\cite{yang2025qwen25} and code-specialized~\cite{hui2025qwen25coder} LLMs over a wide range of sizes. Their scalability, architectural optimizations, and~strong reasoning capabilities make them valuable for benchmarking efficiency and specialization. Here, the~most recent 2.5 versions are tested---namely the specialized coder implementations (0.5B, 1.5B, and~3B) and the general-purpose 0.5B model---as well as QwQ (Qwen with Questions) 32B model with advanced reasoning~capabilities.

\subsection{Implementation}
\label{sec:methods:impl}

\begin{figure}
    \includegraphics[width=0.99\linewidth]{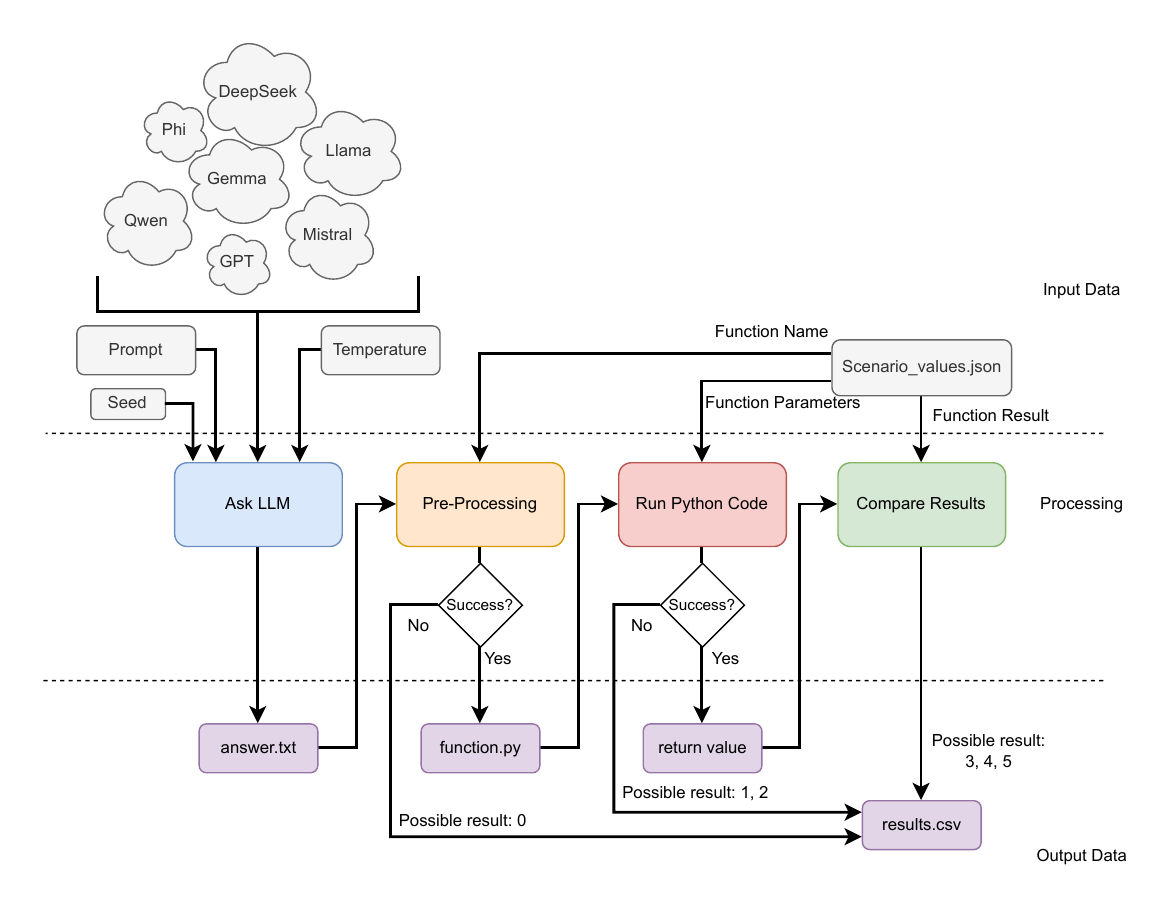}
    \caption{Validation pipeline for the results of LLMs under~study.}
	\label{fig:block_architecture}
\end{figure}

The pipeline for submitting a prompt to an LLM, obtaining a response, extracting a Python function, and~executing it is illustrated in Figure~\ref{fig:block_architecture}.
The process begins by iterating through a predefined set of LLMs, seeds, temperatures, and~prompts. Each prompt is submitted to the corresponding LLM, and~its response is stored in a text file. Next, the~function from each stored response is extracted by searching for the function definition (e.g., `\texttt{def requested\_function():}') and capturing all internal code up to the last properly indented `\texttt{return}' statement. This ensures that functions defined within the external function do not prematurely terminate the extraction. The~extracted function is then recorded in a Python file for execution. If~the function is not successfully extracted---such as when the defined function name does not match the expected one---this information is logged in the results file, and~a score of zero is assigned for that LLM, seed, temperature, and~prompt~combination.

If the Python function is correctly generated and extracted, it is tested under Python~3.9.6 using the data provided for each scenario (presented in Section~\ref{sec:methods:expsetup}). One of three possible outcomes may occur:

\begin{itemize}
    \item The code contains a syntax error and does not compile, in~which case a score of $1$ is recorded in the results file;
    \item The code executes but encounters a runtime error, resulting in an exception, in~which case a score of $2$ is stored in the results file;
    \item The code executes successfully and returns a result, in~which case the score ranges from $3$ to $5$, as~detailed below.
\end{itemize}

If the code executes successfully, the~function's output is evaluated as follows: if the returned value is of a different type than expected (e.g., a~\texttt{float} instead of an \texttt{int}), a~score of $3$ is recorded in the results file. This type check is performed broadly; for example, if~an integer is expected, types such as \texttt{int}, \texttt{np.int32}, or~\texttt{np.int64} are considered valid (where \texttt{np} refers to the NumPy library). If~the type is correct, the~next step is to verify whether the returned value matches the expected value. For~floating-point comparisons, a~tolerance of $1\%$ is allowed. If~the result is incorrect, a~score of $4$ is assigned. Finally, if~the returned value is correct, a~score of $5$ is recorded, indicating 100\% functionally correct code. At~the end of this process, a~file containing all recorded scores is available for~analysis.

In summary, scores between $0$ and $5$ are characterized as follows:

\begin{enumerate}\addtocounter{enumi}{-1}
    \item No Python file was generated---This indicates that the LLM did not generate a Python function or that the generated function does not have the name specified in the prompt.
    \item Syntax error---The code does not compile.
    \item Runtime error---The code is valid Python but has logic incongruencies and/or does not conform to the prompt requirements.
    \item Code runs but returns an incorrect data type---For Prompts 1 and 2, it should return an integer (the index value), while in Prompt~3, it should return a float.
    \item Code runs but returns an incorrect result.
    \item Code runs and returns the correct result.
\end{enumerate}

\subsection{Experimental~Setup}
\label{sec:methods:expsetup}

To thoroughly test the capabilities of the models listed in Section~\ref{sec:methods:llms}, the~prompts presented in Section~\ref{sec:methods:scenarios} were individually submitted to LLMs using six different pseudo-random number generator seeds across six temperature values, in~a total of 36 submissions per prompt for each LLM. Temperatures were increased in \num{0.2} increments from \num{0.0} to \num{1.0} for locally executed LLMs via Ollama. Although~Ollama accepts temperatures in the range of \num{0.0}--\num{1.0}, both DeepSeek-V3 and GPT-4, executed through their online APIs, accept temperatures in the  \num{0.0}--\num{2.0} range. Therefore, temperatures were doubled for these models. For~example, and~for the purpose of this study, a~temperature of \num{0.6} in local models is doubled to \num{1.2} when submitting a prompt to online~LLMs.

The LLM-generated Python functions were tested with the following input data, and return values for each prompt were expected:

\begin{description}
    \item[\textbf{Prompt~1}] The input data are an array of four positions, namely $[(2,5), (7,7), (1,8), (1, 0.5)]$. The~expected return value is $3$, corresponding to coordinate $(1, 0.5)$, which is the closest one to the BS, which is fixed at $(0,0)$.
    \item[\textbf{Prompt~2}] The input data are an array containing the following coordinates:
        \begin{center}
        \begin{tabular}{lcl}
            BS / UAV    & $\rightarrow$ & $(38.759297963817374, -9.154483012234662)$ \\
            Position 1  & $\rightarrow$ & $(38.749330295687805, -9.15304293547367)$  \\ 
            Position 2  & $\rightarrow$ & $(38.75727072916799, -9.157797377555926)$  \\ 
            Position 3  & $\rightarrow$ & $(38.737648166512336, -9.138660615310467)$ \\ 
            Position 4  & $\rightarrow$ & $(38.76841010033327, -9.160013961052972)$  \\ 
        \end{tabular}
        \end{center}
        The expected return value is $2$, corresponding to the index of Position 2, which minimizes the power loss.
    \item[\textbf{Prompt~3}] The input data are the same as in Prompt~2, but~the expected value is   $-50.33$ dBm, which is the minimal loss, obtained at Position 2.
\end{description}

As described in Section~\ref{sec:methods:impl}, the~capabilities of the different LLMs in correctly answering Prompts 1--3 are assessed using a score between 0 and 5. For~six submissions (one per seed) for each prompt--model--temperature combination, four summary statistics are calculated and presented: the mean score, a~non-parametric 95\% confidence interval around the mean, the~percentage of perfect scores (score equal to 5), and~a histogram of score distribution. These metrics allow for a detailed performance investigation of the capabilities of the 16~tested models to generate Python code to solve the three progressively complex LoRaWAN-related~prompts.

In addition to these summary statistics, a~formal statistical comparison between models is conducted using stratified permutation tests~\cite{good2004permutation}. To~account for varying prompt difficulty, model performance is stratified by prompt, allowing all three prompts to be included in a unified testing procedure. For~each pairwise comparison between two models at a given temperature, scores are pooled by prompt (six scores per model per prompt, 12 in total), and~a one-sided permutation test is applied. The~test statistic is the sum of mean rank differences across prompts. All $\binom{12}{6} = 924$ possible permutations of model labels are precomputed per prompt, and~1000 stratified permutations are generated by randomly selecting one permutation per prompt and combining them. The~resulting null distribution is used to estimate the probability of obtaining a test statistic as large or larger than the observed one under the null hypothesis of no difference. The~tests are one-sided, since the goal is to determine whether one model significantly outperforms another---not whether it is worse. Finally, multiple testing correction is applied using the Benjamini--Hochberg procedure to control the false discovery rate (FDR) across all comparisons~\cite{benjamini1995controlling}.

\section{Results}
\label{sec:results}

Results for the simpler Prompt~1 are shown in Figure~\ref{fig:res_prompt1} and Table~\ref{tab:res_prompt1}. While all models generated accurate code for certain seed/temperature combinations, DeepSeek-V3 and Phi-4 stood out, consistently providing correct answers across all seeds and temperatures. The~three LLaMA models, the~three Qwen coder models, and~GPT-4 also demonstrated strong performance, reliably generating correct code for at least a subset of temperature values---typically at lower settings. Interestingly, GPT-4 exhibited a significant drop in answer quality at temperatures of $1.6$ and higher (i.e., $2 \times 0.8$), with~responses becoming essentially random at the highest temperature. In~contrast, the~DeepSeek-R1 models (7B and 70B), the~Gemma models (2B), the~Mistral model (7B), and~the non-coder Qwen models (2.5-0.5B and QwQ-32B) failed to consistently produce correct~answers.

Results for the slightly more complex Prompt~2, for~which the UAV position is given as a function argument (i.e., it is not predefined within the function) and actual geographical coordinates are used, are shown in Figure~\ref{fig:res_prompt2} and Table~\ref{tab:res_prompt2}. Only four models consistently generated accurate code: the larger online DeepSeek-V3 and GPT-4 models, as~well as the smaller, locally tested LLaMA-3.3 and Phi-4. However, the~drop in performance for GPT-4 at higher temperatures is even more pronounced for this prompt. Conversely, Gemma (2B), Mistral (7B), and~both 0.5B Qwen models failed to produce a single correct~answer.

\newcommand{\histw}{0.9cm}

\newcommand{\figurecaption}[1]{%
Prompt #1 mean answer score for the tested models over several temperatures. Each combination of model and temperature was tested with 6 different seeds. Error bars denote a 95\% confidence interval. Temperatures for online models, \texttt{deepseek-v3} and \texttt{gpt-4-0613}, are twice the displayed values.
}

\newcommand{\tablecaption}[1]{%
Prompt #1 answer statistics, namely the percentage of correct answers (score equal to 5) and histogram of scores (0--5) for the tested models over several temperatures. Each combination of model and temperature was tested with 6 different seeds. Temperatures for online models, \texttt{deepseek-v3} and \texttt{gpt-4-0613}, are twice the displayed values.
}

\afterpage{

\begin{figure}[!h]
    \centering
    \resizebox{\textwidth}{!}{  
    \includegraphics[width=1\linewidth]{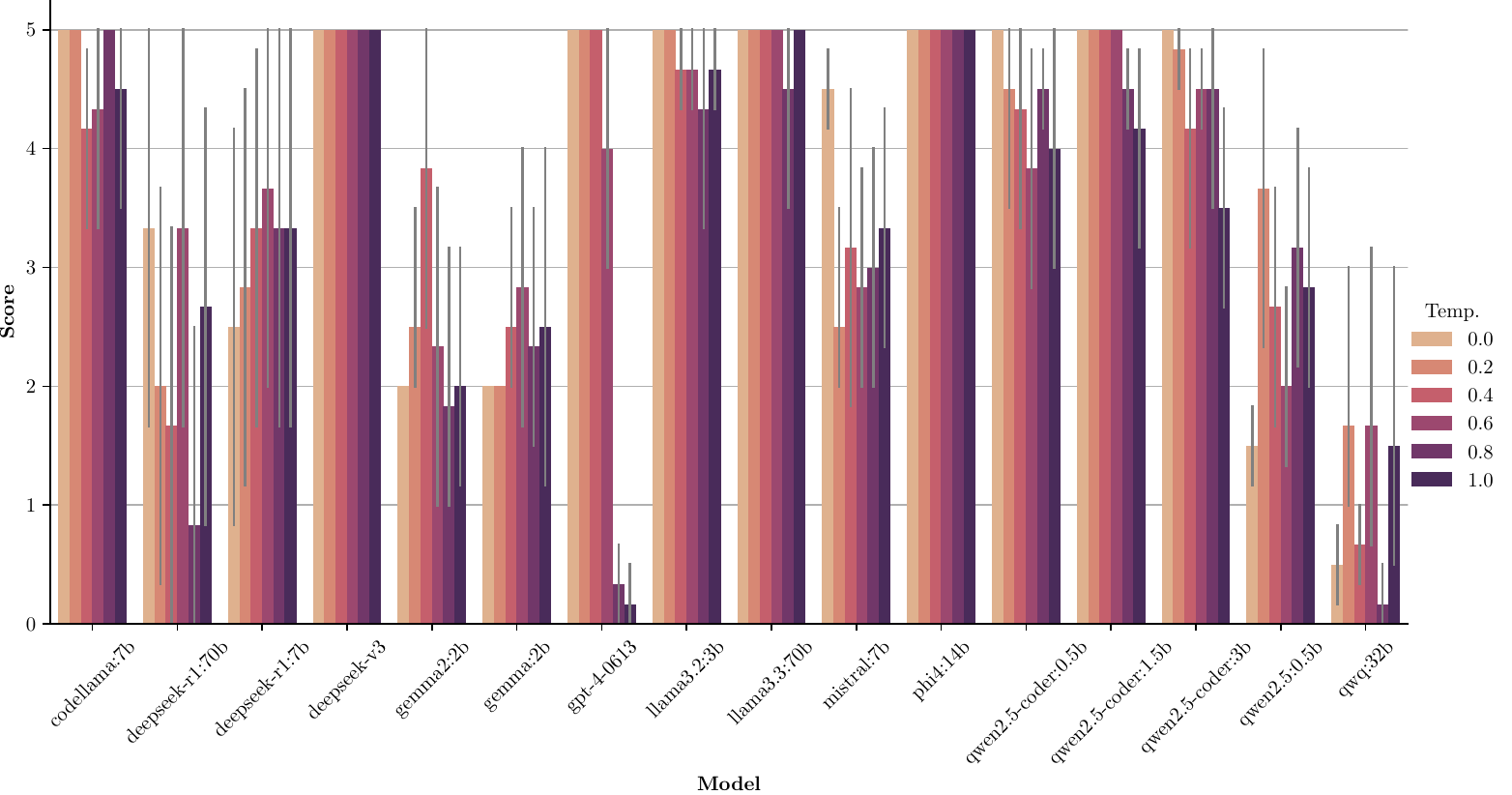}
    }
    \caption{\figurecaption{1}}
	\label{fig:res_prompt1}
\end{figure}

\begin{table}[!h]
    \caption{\tablecaption{1}} 
    \label{tab:res_prompt1}
    \resizebox{\textwidth}{!}{  
    \begin{tabular}{lr@{}rr@{}rr@{}rr@{}rr@{}rr@{}rr@{}r}
        \toprule
        \multirow{2.2}{*}{Model} & \multicolumn{14}{l}{Temperature} \\
              \cmidrule(l){2-15}
              & \multicolumn{2}{r}{0.0} & \multicolumn{2}{r}{0.2} & \multicolumn{2}{r}{0.4} & \multicolumn{2}{r}{0.6} & \multicolumn{2}{r}{0.8} & \multicolumn{2}{r}{1.0} & \multicolumn{2}{r}{Overall} \\
        \midrule

{\small\texttt{codellama:7b}}
& {\footnotesize\num{ 100.0}\%} & \resizebox{\histw}{!}{\drawHistogram{0}{0}{0}{0}{0}{6}}
& {\footnotesize\num{ 100.0}\%} & \resizebox{\histw}{!}{\drawHistogram{0}{0}{0}{0}{0}{6}}
& {\footnotesize\num{ 50.0}\%} & \resizebox{\histw}{!}{\drawHistogram{0}{0}{1}{0}{2}{3}}
& {\footnotesize\num{ 66.66666666666666}\%} & \resizebox{\histw}{!}{\drawHistogram{0}{0}{1}{0}{1}{4}}
& {\footnotesize\num{ 100.0}\%} & \resizebox{\histw}{!}{\drawHistogram{0}{0}{0}{0}{0}{6}}
& {\footnotesize\num{ 83.33333333333334}\%} & \resizebox{\histw}{!}{\drawHistogram{0}{0}{1}{0}{0}{5}}
& {\footnotesize\num{ 83.33333333333334}\%} & \resizebox{\histw}{!}{\drawHistogram{0}{0}{3}{0}{3}{30}}
\\
{\small\texttt{deepseek-r1:70b}}
& {\footnotesize\num{ 66.66666666666666}\%} & \resizebox{\histw}{!}{\drawHistogram{2}{0}{0}{0}{0}{4}}
& {\footnotesize\num{ 33.33333333333333}\%} & \resizebox{\histw}{!}{\drawHistogram{2}{2}{0}{0}{0}{2}}
& {\footnotesize\num{ 33.33333333333333}\%} & \resizebox{\histw}{!}{\drawHistogram{4}{0}{0}{0}{0}{2}}
& {\footnotesize\num{ 66.66666666666666}\%} & \resizebox{\histw}{!}{\drawHistogram{2}{0}{0}{0}{0}{4}}
& {\footnotesize\num{ 16.666666666666664}\%} & \resizebox{\histw}{!}{\drawHistogram{5}{0}{0}{0}{0}{1}}
& {\footnotesize\num{ 50.0}\%} & \resizebox{\histw}{!}{\drawHistogram{2}{1}{0}{0}{0}{3}}
& {\footnotesize\num{ 44.44444444444444}\%} & \resizebox{\histw}{!}{\drawHistogram{17}{3}{0}{0}{0}{16}}
\\
{\small\texttt{deepseek-r1:7b}}
& {\footnotesize\num{ 50.0}\%} & \resizebox{\histw}{!}{\drawHistogram{3}{0}{0}{0}{0}{3}}
& {\footnotesize\num{ 50.0}\%} & \resizebox{\histw}{!}{\drawHistogram{2}{0}{1}{0}{0}{3}}
& {\footnotesize\num{ 50.0}\%} & \resizebox{\histw}{!}{\drawHistogram{1}{1}{0}{0}{1}{3}}
& {\footnotesize\num{ 66.66666666666666}\%} & \resizebox{\histw}{!}{\drawHistogram{1}{0}{1}{0}{0}{4}}
& {\footnotesize\num{ 66.66666666666666}\%} & \resizebox{\histw}{!}{\drawHistogram{2}{0}{0}{0}{0}{4}}
& {\footnotesize\num{ 66.66666666666666}\%} & \resizebox{\histw}{!}{\drawHistogram{2}{0}{0}{0}{0}{4}}
& {\footnotesize\num{ 58.333333333333336}\%} & \resizebox{\histw}{!}{\drawHistogram{11}{1}{2}{0}{1}{21}}
\\
{\small\texttt{deepseek-v3}}
& {\footnotesize\num{ 100.0}\%} & \resizebox{\histw}{!}{\drawHistogram{0}{0}{0}{0}{0}{6}}
& {\footnotesize\num{ 100.0}\%} & \resizebox{\histw}{!}{\drawHistogram{0}{0}{0}{0}{0}{6}}
& {\footnotesize\num{ 100.0}\%} & \resizebox{\histw}{!}{\drawHistogram{0}{0}{0}{0}{0}{6}}
& {\footnotesize\num{ 100.0}\%} & \resizebox{\histw}{!}{\drawHistogram{0}{0}{0}{0}{0}{6}}
& {\footnotesize\num{ 100.0}\%} & \resizebox{\histw}{!}{\drawHistogram{0}{0}{0}{0}{0}{6}}
& {\footnotesize\num{ 100.0}\%} & \resizebox{\histw}{!}{\drawHistogram{0}{0}{0}{0}{0}{6}}
& {\footnotesize\num{ 100.0}\%} & \resizebox{\histw}{!}{\drawHistogram{0}{0}{0}{0}{0}{36}}
\\
{\small\texttt{gemma2:2b}}
& {\footnotesize\num{ 0.0}\%} & \resizebox{\histw}{!}{\drawHistogram{0}{0}{6}{0}{0}{0}}
& {\footnotesize\num{ 16.666666666666664}\%} & \resizebox{\histw}{!}{\drawHistogram{0}{0}{5}{0}{0}{1}}
& {\footnotesize\num{ 66.66666666666666}\%} & \resizebox{\histw}{!}{\drawHistogram{0}{1}{1}{0}{0}{4}}
& {\footnotesize\num{ 33.33333333333333}\%} & \resizebox{\histw}{!}{\drawHistogram{0}{4}{0}{0}{0}{2}}
& {\footnotesize\num{ 16.666666666666664}\%} & \resizebox{\histw}{!}{\drawHistogram{0}{4}{1}{0}{0}{1}}
& {\footnotesize\num{ 16.666666666666664}\%} & \resizebox{\histw}{!}{\drawHistogram{0}{3}{2}{0}{0}{1}}
& {\footnotesize\num{ 25.0}\%} & \resizebox{\histw}{!}{\drawHistogram{0}{12}{15}{0}{0}{9}}
\\
{\small\texttt{gemma:2b}}
& {\footnotesize\num{ 0.0}\%} & \resizebox{\histw}{!}{\drawHistogram{0}{0}{6}{0}{0}{0}}
& {\footnotesize\num{ 0.0}\%} & \resizebox{\histw}{!}{\drawHistogram{0}{0}{6}{0}{0}{0}}
& {\footnotesize\num{ 16.666666666666664}\%} & \resizebox{\histw}{!}{\drawHistogram{0}{0}{5}{0}{0}{1}}
& {\footnotesize\num{ 33.33333333333333}\%} & \resizebox{\histw}{!}{\drawHistogram{0}{1}{3}{0}{0}{2}}
& {\footnotesize\num{ 16.666666666666664}\%} & \resizebox{\histw}{!}{\drawHistogram{0}{1}{4}{0}{0}{1}}
& {\footnotesize\num{ 33.33333333333333}\%} & \resizebox{\histw}{!}{\drawHistogram{0}{3}{1}{0}{0}{2}}
& {\footnotesize\num{ 16.666666666666664}\%} & \resizebox{\histw}{!}{\drawHistogram{0}{5}{25}{0}{0}{6}}
\\
{\small\texttt{gpt-4-0613}}
& {\footnotesize\num{ 100.0}\%} & \resizebox{\histw}{!}{\drawHistogram{0}{0}{0}{0}{0}{6}}
& {\footnotesize\num{ 100.0}\%} & \resizebox{\histw}{!}{\drawHistogram{0}{0}{0}{0}{0}{6}}
& {\footnotesize\num{ 100.0}\%} & \resizebox{\histw}{!}{\drawHistogram{0}{0}{0}{0}{0}{6}}
& {\footnotesize\num{ 66.66666666666666}\%} & \resizebox{\histw}{!}{\drawHistogram{0}{0}{2}{0}{0}{4}}
& {\footnotesize\num{ 0.0}\%} & \resizebox{\histw}{!}{\drawHistogram{4}{2}{0}{0}{0}{0}}
& {\footnotesize\num{ 0.0}\%} & \resizebox{\histw}{!}{\drawHistogram{5}{1}{0}{0}{0}{0}}
& {\footnotesize\num{ 61.111111111111114}\%} & \resizebox{\histw}{!}{\drawHistogram{9}{3}{2}{0}{0}{22}}
\\
{\small\texttt{llama3.2:3b}}
& {\footnotesize\num{ 100.0}\%} & \resizebox{\histw}{!}{\drawHistogram{0}{0}{0}{0}{0}{6}}
& {\footnotesize\num{ 100.0}\%} & \resizebox{\histw}{!}{\drawHistogram{0}{0}{0}{0}{0}{6}}
& {\footnotesize\num{ 66.66666666666666}\%} & \resizebox{\histw}{!}{\drawHistogram{0}{0}{0}{0}{2}{4}}
& {\footnotesize\num{ 66.66666666666666}\%} & \resizebox{\histw}{!}{\drawHistogram{0}{0}{0}{0}{2}{4}}
& {\footnotesize\num{ 66.66666666666666}\%} & \resizebox{\histw}{!}{\drawHistogram{0}{0}{1}{0}{1}{4}}
& {\footnotesize\num{ 66.66666666666666}\%} & \resizebox{\histw}{!}{\drawHistogram{0}{0}{0}{0}{2}{4}}
& {\footnotesize\num{ 77.77777777777779}\%} & \resizebox{\histw}{!}{\drawHistogram{0}{0}{1}{0}{7}{28}}
\\
{\small\texttt{llama3.3:70b}}
& {\footnotesize\num{ 100.0}\%} & \resizebox{\histw}{!}{\drawHistogram{0}{0}{0}{0}{0}{6}}
& {\footnotesize\num{ 100.0}\%} & \resizebox{\histw}{!}{\drawHistogram{0}{0}{0}{0}{0}{6}}
& {\footnotesize\num{ 100.0}\%} & \resizebox{\histw}{!}{\drawHistogram{0}{0}{0}{0}{0}{6}}
& {\footnotesize\num{ 100.0}\%} & \resizebox{\histw}{!}{\drawHistogram{0}{0}{0}{0}{0}{6}}
& {\footnotesize\num{ 83.33333333333334}\%} & \resizebox{\histw}{!}{\drawHistogram{0}{0}{1}{0}{0}{5}}
& {\footnotesize\num{ 100.0}\%} & \resizebox{\histw}{!}{\drawHistogram{0}{0}{0}{0}{0}{6}}
& {\footnotesize\num{ 97.22222222222221}\%} & \resizebox{\histw}{!}{\drawHistogram{0}{0}{1}{0}{0}{35}}
\\
{\small\texttt{mistral:7b}}
& {\footnotesize\num{ 50.0}\%} & \resizebox{\histw}{!}{\drawHistogram{0}{0}{0}{0}{3}{3}}
& {\footnotesize\num{ 16.666666666666664}\%} & \resizebox{\histw}{!}{\drawHistogram{0}{0}{5}{0}{0}{1}}
& {\footnotesize\num{ 50.0}\%} & \resizebox{\histw}{!}{\drawHistogram{0}{2}{1}{0}{0}{3}}
& {\footnotesize\num{ 16.666666666666664}\%} & \resizebox{\histw}{!}{\drawHistogram{0}{0}{4}{0}{1}{1}}
& {\footnotesize\num{ 33.33333333333333}\%} & \resizebox{\histw}{!}{\drawHistogram{0}{0}{4}{0}{0}{2}}
& {\footnotesize\num{ 33.33333333333333}\%} & \resizebox{\histw}{!}{\drawHistogram{0}{0}{3}{0}{1}{2}}
& {\footnotesize\num{ 33.33333333333333}\%} & \resizebox{\histw}{!}{\drawHistogram{0}{2}{17}{0}{5}{12}}
\\
{\small\texttt{phi4:14b}}
& {\footnotesize\num{ 100.0}\%} & \resizebox{\histw}{!}{\drawHistogram{0}{0}{0}{0}{0}{6}}
& {\footnotesize\num{ 100.0}\%} & \resizebox{\histw}{!}{\drawHistogram{0}{0}{0}{0}{0}{6}}
& {\footnotesize\num{ 100.0}\%} & \resizebox{\histw}{!}{\drawHistogram{0}{0}{0}{0}{0}{6}}
& {\footnotesize\num{ 100.0}\%} & \resizebox{\histw}{!}{\drawHistogram{0}{0}{0}{0}{0}{6}}
& {\footnotesize\num{ 100.0}\%} & \resizebox{\histw}{!}{\drawHistogram{0}{0}{0}{0}{0}{6}}
& {\footnotesize\num{ 100.0}\%} & \resizebox{\histw}{!}{\drawHistogram{0}{0}{0}{0}{0}{6}}
& {\footnotesize\num{ 100.0}\%} & \resizebox{\histw}{!}{\drawHistogram{0}{0}{0}{0}{0}{36}}
\\
{\small\texttt{qwen2.5-coder:0.5b}}
& {\footnotesize\num{ 100.0}\%} & \resizebox{\histw}{!}{\drawHistogram{0}{0}{0}{0}{0}{6}}
& {\footnotesize\num{ 83.33333333333334}\%} & \resizebox{\histw}{!}{\drawHistogram{0}{0}{1}{0}{0}{5}}
& {\footnotesize\num{ 66.66666666666666}\%} & \resizebox{\histw}{!}{\drawHistogram{0}{0}{1}{0}{1}{4}}
& {\footnotesize\num{ 50.0}\%} & \resizebox{\histw}{!}{\drawHistogram{0}{0}{2}{0}{1}{3}}
& {\footnotesize\num{ 50.0}\%} & \resizebox{\histw}{!}{\drawHistogram{0}{0}{0}{0}{3}{3}}
& {\footnotesize\num{ 66.66666666666666}\%} & \resizebox{\histw}{!}{\drawHistogram{0}{0}{2}{0}{0}{4}}
& {\footnotesize\num{ 69.44444444444444}\%} & \resizebox{\histw}{!}{\drawHistogram{0}{0}{6}{0}{5}{25}}
\\
{\small\texttt{qwen2.5-coder:1.5b}}
& {\footnotesize\num{ 100.0}\%} & \resizebox{\histw}{!}{\drawHistogram{0}{0}{0}{0}{0}{6}}
& {\footnotesize\num{ 100.0}\%} & \resizebox{\histw}{!}{\drawHistogram{0}{0}{0}{0}{0}{6}}
& {\footnotesize\num{ 100.0}\%} & \resizebox{\histw}{!}{\drawHistogram{0}{0}{0}{0}{0}{6}}
& {\footnotesize\num{ 100.0}\%} & \resizebox{\histw}{!}{\drawHistogram{0}{0}{0}{0}{0}{6}}
& {\footnotesize\num{ 50.0}\%} & \resizebox{\histw}{!}{\drawHistogram{0}{0}{0}{0}{3}{3}}
& {\footnotesize\num{ 50.0}\%} & \resizebox{\histw}{!}{\drawHistogram{0}{0}{1}{0}{2}{3}}
& {\footnotesize\num{ 83.33333333333334}\%} & \resizebox{\histw}{!}{\drawHistogram{0}{0}{1}{0}{5}{30}}
\\
{\small\texttt{qwen2.5-coder:3b}}
& {\footnotesize\num{ 100.0}\%} & \resizebox{\histw}{!}{\drawHistogram{0}{0}{0}{0}{0}{6}}
& {\footnotesize\num{ 83.33333333333334}\%} & \resizebox{\histw}{!}{\drawHistogram{0}{0}{0}{0}{1}{5}}
& {\footnotesize\num{ 50.0}\%} & \resizebox{\histw}{!}{\drawHistogram{0}{0}{1}{0}{2}{3}}
& {\footnotesize\num{ 50.0}\%} & \resizebox{\histw}{!}{\drawHistogram{0}{0}{0}{0}{3}{3}}
& {\footnotesize\num{ 83.33333333333334}\%} & \resizebox{\histw}{!}{\drawHistogram{0}{0}{1}{0}{0}{5}}
& {\footnotesize\num{ 16.666666666666664}\%} & \resizebox{\histw}{!}{\drawHistogram{0}{0}{2}{0}{3}{1}}
& {\footnotesize\num{ 63.888888888888886}\%} & \resizebox{\histw}{!}{\drawHistogram{0}{0}{4}{0}{9}{23}}
\\
{\small\texttt{qwen2.5:0.5b}}
& {\footnotesize\num{ 0.0}\%} & \resizebox{\histw}{!}{\drawHistogram{0}{3}{3}{0}{0}{0}}
& {\footnotesize\num{ 50.0}\%} & \resizebox{\histw}{!}{\drawHistogram{0}{1}{1}{0}{1}{3}}
& {\footnotesize\num{ 16.666666666666664}\%} & \resizebox{\histw}{!}{\drawHistogram{0}{1}{3}{0}{1}{1}}
& {\footnotesize\num{ 0.0}\%} & \resizebox{\histw}{!}{\drawHistogram{0}{2}{3}{0}{1}{0}}
& {\footnotesize\num{ 16.666666666666664}\%} & \resizebox{\histw}{!}{\drawHistogram{0}{1}{1}{1}{2}{1}}
& {\footnotesize\num{ 16.666666666666664}\%} & \resizebox{\histw}{!}{\drawHistogram{0}{0}{4}{0}{1}{1}}
& {\footnotesize\num{ 16.666666666666664}\%} & \resizebox{\histw}{!}{\drawHistogram{0}{8}{15}{1}{6}{6}}
\\
{\small\texttt{qwq:32b}}
& {\footnotesize\num{ 0.0}\%} & \resizebox{\histw}{!}{\drawHistogram{3}{3}{0}{0}{0}{0}}
& {\footnotesize\num{ 16.666666666666664}\%} & \resizebox{\histw}{!}{\drawHistogram{0}{5}{0}{0}{0}{1}}
& {\footnotesize\num{ 0.0}\%} & \resizebox{\histw}{!}{\drawHistogram{2}{4}{0}{0}{0}{0}}
& {\footnotesize\num{ 16.666666666666664}\%} & \resizebox{\histw}{!}{\drawHistogram{1}{3}{1}{0}{0}{1}}
& {\footnotesize\num{ 0.0}\%} & \resizebox{\histw}{!}{\drawHistogram{5}{1}{0}{0}{0}{0}}
& {\footnotesize\num{ 16.666666666666664}\%} & \resizebox{\histw}{!}{\drawHistogram{1}{4}{0}{0}{0}{1}}
& {\footnotesize\num{ 8.333333333333332}\%} & \resizebox{\histw}{!}{\drawHistogram{12}{20}{1}{0}{0}{3}}
\\

        \bottomrule
    \end{tabular}
    }
\end{table}

} 

\clearpage

\afterpage{

\clearpage

\begin{figure}[ht!]
    \centering
    \resizebox{\textwidth}{!}{  
    \includegraphics[width=1\linewidth]{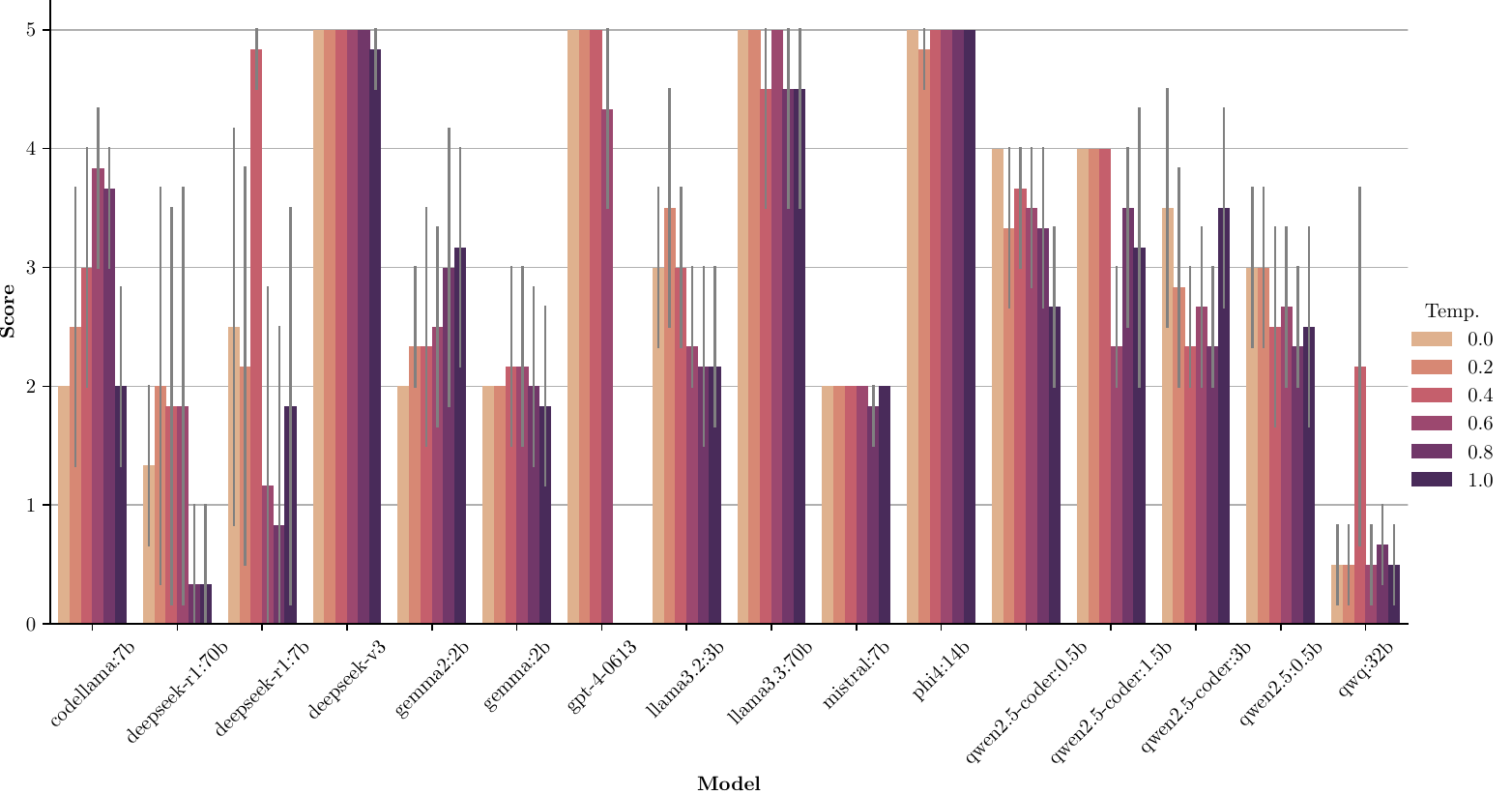}
    }
    \caption{\figurecaption{2}}
	\label{fig:res_prompt2}
\end{figure}

\begin{table}[ht!]
    \caption{\tablecaption{2}} 
    \label{tab:res_prompt2}
    \resizebox{\textwidth}{!}{  
    \begin{tabular}{lr@{}rr@{}rr@{}rr@{}rr@{}rr@{}rr@{}r}
        \toprule
        \multirow{2.2}{*}{Model} & \multicolumn{14}{l}{Temperature} \\
              \cmidrule(l){2-15}
              & \multicolumn{2}{r}{0.0} & \multicolumn{2}{r}{0.2} & \multicolumn{2}{r}{0.4} & \multicolumn{2}{r}{0.6} & \multicolumn{2}{r}{0.8} & \multicolumn{2}{r}{1.0} & \multicolumn{2}{r}{Overall} \\
        \midrule

{\small\texttt{codellama:7b}}
& {\footnotesize\num{ 0.0}\%} & \resizebox{\histw}{!}{\drawHistogram{0}{0}{6}{0}{0}{0}}
& {\footnotesize\num{ 16.666666666666664}\%} & \resizebox{\histw}{!}{\drawHistogram{0}{2}{2}{0}{1}{1}}
& {\footnotesize\num{ 16.666666666666664}\%} & \resizebox{\histw}{!}{\drawHistogram{0}{1}{2}{0}{2}{1}}
& {\footnotesize\num{ 16.666666666666664}\%} & \resizebox{\histw}{!}{\drawHistogram{0}{0}{1}{0}{4}{1}}
& {\footnotesize\num{ 0.0}\%} & \resizebox{\histw}{!}{\drawHistogram{0}{0}{1}{0}{5}{0}}
& {\footnotesize\num{ 0.0}\%} & \resizebox{\histw}{!}{\drawHistogram{0}{2}{3}{0}{1}{0}}
& {\footnotesize\num{ 8.333333333333332}\%} & \resizebox{\histw}{!}{\drawHistogram{0}{5}{15}{0}{13}{3}}
\\
{\small\texttt{deepseek-r1:70b}}
& {\footnotesize\num{ 0.0}\%} & \resizebox{\histw}{!}{\drawHistogram{2}{0}{4}{0}{0}{0}}
& {\footnotesize\num{ 33.33333333333333}\%} & \resizebox{\histw}{!}{\drawHistogram{3}{0}{1}{0}{0}{2}}
& {\footnotesize\num{ 33.33333333333333}\%} & \resizebox{\histw}{!}{\drawHistogram{3}{1}{0}{0}{0}{2}}
& {\footnotesize\num{ 33.33333333333333}\%} & \resizebox{\histw}{!}{\drawHistogram{3}{1}{0}{0}{0}{2}}
& {\footnotesize\num{ 0.0}\%} & \resizebox{\histw}{!}{\drawHistogram{5}{0}{1}{0}{0}{0}}
& {\footnotesize\num{ 0.0}\%} & \resizebox{\histw}{!}{\drawHistogram{5}{0}{1}{0}{0}{0}}
& {\footnotesize\num{ 16.666666666666664}\%} & \resizebox{\histw}{!}{\drawHistogram{21}{2}{7}{0}{0}{6}}
\\
{\small\texttt{deepseek-r1:7b}}
& {\footnotesize\num{ 50.0}\%} & \resizebox{\histw}{!}{\drawHistogram{3}{0}{0}{0}{0}{3}}
& {\footnotesize\num{ 33.33333333333333}\%} & \resizebox{\histw}{!}{\drawHistogram{2}{1}{1}{0}{0}{2}}
& {\footnotesize\num{ 83.33333333333334}\%} & \resizebox{\histw}{!}{\drawHistogram{0}{0}{0}{0}{1}{5}}
& {\footnotesize\num{ 16.666666666666664}\%} & \resizebox{\histw}{!}{\drawHistogram{4}{0}{1}{0}{0}{1}}
& {\footnotesize\num{ 16.666666666666664}\%} & \resizebox{\histw}{!}{\drawHistogram{5}{0}{0}{0}{0}{1}}
& {\footnotesize\num{ 33.33333333333333}\%} & \resizebox{\histw}{!}{\drawHistogram{3}{1}{0}{0}{0}{2}}
& {\footnotesize\num{ 38.88888888888889}\%} & \resizebox{\histw}{!}{\drawHistogram{17}{2}{2}{0}{1}{14}}
\\
{\small\texttt{deepseek-v3}}
& {\footnotesize\num{ 100.0}\%} & \resizebox{\histw}{!}{\drawHistogram{0}{0}{0}{0}{0}{6}}
& {\footnotesize\num{ 100.0}\%} & \resizebox{\histw}{!}{\drawHistogram{0}{0}{0}{0}{0}{6}}
& {\footnotesize\num{ 100.0}\%} & \resizebox{\histw}{!}{\drawHistogram{0}{0}{0}{0}{0}{6}}
& {\footnotesize\num{ 100.0}\%} & \resizebox{\histw}{!}{\drawHistogram{0}{0}{0}{0}{0}{6}}
& {\footnotesize\num{ 100.0}\%} & \resizebox{\histw}{!}{\drawHistogram{0}{0}{0}{0}{0}{6}}
& {\footnotesize\num{ 83.33333333333334}\%} & \resizebox{\histw}{!}{\drawHistogram{0}{0}{0}{0}{1}{5}}
& {\footnotesize\num{ 97.22222222222221}\%} & \resizebox{\histw}{!}{\drawHistogram{0}{0}{0}{0}{1}{35}}
\\
{\small\texttt{gemma2:2b}}
& {\footnotesize\num{ 0.0}\%} & \resizebox{\histw}{!}{\drawHistogram{0}{0}{6}{0}{0}{0}}
& {\footnotesize\num{ 0.0}\%} & \resizebox{\histw}{!}{\drawHistogram{0}{0}{5}{0}{1}{0}}
& {\footnotesize\num{ 16.666666666666664}\%} & \resizebox{\histw}{!}{\drawHistogram{0}{1}{4}{0}{0}{1}}
& {\footnotesize\num{ 0.0}\%} & \resizebox{\histw}{!}{\drawHistogram{0}{1}{3}{0}{2}{0}}
& {\footnotesize\num{ 16.666666666666664}\%} & \resizebox{\histw}{!}{\drawHistogram{0}{1}{2}{0}{2}{1}}
& {\footnotesize\num{ 0.0}\%} & \resizebox{\histw}{!}{\drawHistogram{0}{1}{1}{0}{4}{0}}
& {\footnotesize\num{ 5.555555555555555}\%} & \resizebox{\histw}{!}{\drawHistogram{0}{4}{21}{0}{9}{2}}
\\
{\small\texttt{gemma:2b}}
& {\footnotesize\num{ 0.0}\%} & \resizebox{\histw}{!}{\drawHistogram{0}{0}{6}{0}{0}{0}}
& {\footnotesize\num{ 0.0}\%} & \resizebox{\histw}{!}{\drawHistogram{0}{0}{6}{0}{0}{0}}
& {\footnotesize\num{ 0.0}\%} & \resizebox{\histw}{!}{\drawHistogram{0}{1}{4}{0}{1}{0}}
& {\footnotesize\num{ 0.0}\%} & \resizebox{\histw}{!}{\drawHistogram{0}{1}{4}{0}{1}{0}}
& {\footnotesize\num{ 0.0}\%} & \resizebox{\histw}{!}{\drawHistogram{0}{2}{3}{0}{1}{0}}
& {\footnotesize\num{ 0.0}\%} & \resizebox{\histw}{!}{\drawHistogram{0}{3}{2}{0}{1}{0}}
& {\footnotesize\num{ 0.0}\%} & \resizebox{\histw}{!}{\drawHistogram{0}{7}{25}{0}{4}{0}}
\\
{\small\texttt{gpt-4-0613}}
& {\footnotesize\num{ 100.0}\%} & \resizebox{\histw}{!}{\drawHistogram{0}{0}{0}{0}{0}{6}}
& {\footnotesize\num{ 100.0}\%} & \resizebox{\histw}{!}{\drawHistogram{0}{0}{0}{0}{0}{6}}
& {\footnotesize\num{ 100.0}\%} & \resizebox{\histw}{!}{\drawHistogram{0}{0}{0}{0}{0}{6}}
& {\footnotesize\num{ 66.66666666666666}\%} & \resizebox{\histw}{!}{\drawHistogram{0}{0}{1}{0}{1}{4}}
& {\footnotesize\num{ 0.0}\%} & \resizebox{\histw}{!}{\drawHistogram{6}{0}{0}{0}{0}{0}}
& {\footnotesize\num{ 0.0}\%} & \resizebox{\histw}{!}{\drawHistogram{6}{0}{0}{0}{0}{0}}
& {\footnotesize\num{ 61.111111111111114}\%} & \resizebox{\histw}{!}{\drawHistogram{12}{0}{1}{0}{1}{22}}
\\
{\small\texttt{llama3.2:3b}}
& {\footnotesize\num{ 0.0}\%} & \resizebox{\histw}{!}{\drawHistogram{0}{0}{3}{0}{3}{0}}
& {\footnotesize\num{ 50.0}\%} & \resizebox{\histw}{!}{\drawHistogram{0}{0}{3}{0}{0}{3}}
& {\footnotesize\num{ 0.0}\%} & \resizebox{\histw}{!}{\drawHistogram{0}{0}{3}{0}{3}{0}}
& {\footnotesize\num{ 0.0}\%} & \resizebox{\histw}{!}{\drawHistogram{0}{0}{5}{0}{1}{0}}
& {\footnotesize\num{ 0.0}\%} & \resizebox{\histw}{!}{\drawHistogram{0}{1}{4}{0}{1}{0}}
& {\footnotesize\num{ 0.0}\%} & \resizebox{\histw}{!}{\drawHistogram{0}{1}{4}{0}{1}{0}}
& {\footnotesize\num{ 8.333333333333332}\%} & \resizebox{\histw}{!}{\drawHistogram{0}{2}{22}{0}{9}{3}}
\\
{\small\texttt{llama3.3:70b}}
& {\footnotesize\num{ 100.0}\%} & \resizebox{\histw}{!}{\drawHistogram{0}{0}{0}{0}{0}{6}}
& {\footnotesize\num{ 100.0}\%} & \resizebox{\histw}{!}{\drawHistogram{0}{0}{0}{0}{0}{6}}
& {\footnotesize\num{ 83.33333333333334}\%} & \resizebox{\histw}{!}{\drawHistogram{0}{0}{1}{0}{0}{5}}
& {\footnotesize\num{ 100.0}\%} & \resizebox{\histw}{!}{\drawHistogram{0}{0}{0}{0}{0}{6}}
& {\footnotesize\num{ 83.33333333333334}\%} & \resizebox{\histw}{!}{\drawHistogram{0}{0}{1}{0}{0}{5}}
& {\footnotesize\num{ 83.33333333333334}\%} & \resizebox{\histw}{!}{\drawHistogram{0}{0}{1}{0}{0}{5}}
& {\footnotesize\num{ 91.66666666666666}\%} & \resizebox{\histw}{!}{\drawHistogram{0}{0}{3}{0}{0}{33}}
\\
{\small\texttt{mistral:7b}}
& {\footnotesize\num{ 0.0}\%} & \resizebox{\histw}{!}{\drawHistogram{0}{0}{6}{0}{0}{0}}
& {\footnotesize\num{ 0.0}\%} & \resizebox{\histw}{!}{\drawHistogram{0}{0}{6}{0}{0}{0}}
& {\footnotesize\num{ 0.0}\%} & \resizebox{\histw}{!}{\drawHistogram{0}{0}{6}{0}{0}{0}}
& {\footnotesize\num{ 0.0}\%} & \resizebox{\histw}{!}{\drawHistogram{0}{0}{6}{0}{0}{0}}
& {\footnotesize\num{ 0.0}\%} & \resizebox{\histw}{!}{\drawHistogram{0}{1}{5}{0}{0}{0}}
& {\footnotesize\num{ 0.0}\%} & \resizebox{\histw}{!}{\drawHistogram{0}{0}{6}{0}{0}{0}}
& {\footnotesize\num{ 0.0}\%} & \resizebox{\histw}{!}{\drawHistogram{0}{1}{35}{0}{0}{0}}
\\
{\small\texttt{phi4:14b}}
& {\footnotesize\num{ 100.0}\%} & \resizebox{\histw}{!}{\drawHistogram{0}{0}{0}{0}{0}{6}}
& {\footnotesize\num{ 83.33333333333334}\%} & \resizebox{\histw}{!}{\drawHistogram{0}{0}{0}{0}{1}{5}}
& {\footnotesize\num{ 100.0}\%} & \resizebox{\histw}{!}{\drawHistogram{0}{0}{0}{0}{0}{6}}
& {\footnotesize\num{ 100.0}\%} & \resizebox{\histw}{!}{\drawHistogram{0}{0}{0}{0}{0}{6}}
& {\footnotesize\num{ 100.0}\%} & \resizebox{\histw}{!}{\drawHistogram{0}{0}{0}{0}{0}{6}}
& {\footnotesize\num{ 100.0}\%} & \resizebox{\histw}{!}{\drawHistogram{0}{0}{0}{0}{0}{6}}
& {\footnotesize\num{ 97.22222222222221}\%} & \resizebox{\histw}{!}{\drawHistogram{0}{0}{0}{0}{1}{35}}
\\
{\small\texttt{qwen2.5-coder:0.5b}}
& {\footnotesize\num{ 0.0}\%} & \resizebox{\histw}{!}{\drawHistogram{0}{0}{0}{0}{6}{0}}
& {\footnotesize\num{ 0.0}\%} & \resizebox{\histw}{!}{\drawHistogram{0}{0}{2}{0}{4}{0}}
& {\footnotesize\num{ 0.0}\%} & \resizebox{\histw}{!}{\drawHistogram{0}{0}{1}{0}{5}{0}}
& {\footnotesize\num{ 0.0}\%} & \resizebox{\histw}{!}{\drawHistogram{0}{0}{1}{1}{4}{0}}
& {\footnotesize\num{ 0.0}\%} & \resizebox{\histw}{!}{\drawHistogram{0}{0}{2}{0}{4}{0}}
& {\footnotesize\num{ 0.0}\%} & \resizebox{\histw}{!}{\drawHistogram{0}{0}{4}{0}{2}{0}}
& {\footnotesize\num{ 0.0}\%} & \resizebox{\histw}{!}{\drawHistogram{0}{0}{10}{1}{25}{0}}
\\
{\small\texttt{qwen2.5-coder:1.5b}}
& {\footnotesize\num{ 0.0}\%} & \resizebox{\histw}{!}{\drawHistogram{0}{0}{0}{0}{6}{0}}
& {\footnotesize\num{ 0.0}\%} & \resizebox{\histw}{!}{\drawHistogram{0}{0}{0}{0}{6}{0}}
& {\footnotesize\num{ 0.0}\%} & \resizebox{\histw}{!}{\drawHistogram{0}{0}{0}{0}{6}{0}}
& {\footnotesize\num{ 0.0}\%} & \resizebox{\histw}{!}{\drawHistogram{0}{0}{5}{0}{1}{0}}
& {\footnotesize\num{ 0.0}\%} & \resizebox{\histw}{!}{\drawHistogram{0}{1}{0}{0}{5}{0}}
& {\footnotesize\num{ 33.33333333333333}\%} & \resizebox{\histw}{!}{\drawHistogram{0}{1}{2}{0}{1}{2}}
& {\footnotesize\num{ 5.555555555555555}\%} & \resizebox{\histw}{!}{\drawHistogram{0}{2}{7}{0}{25}{2}}
\\
{\small\texttt{qwen2.5-coder:3b}}
& {\footnotesize\num{ 50.0}\%} & \resizebox{\histw}{!}{\drawHistogram{0}{0}{3}{0}{0}{3}}
& {\footnotesize\num{ 16.666666666666664}\%} & \resizebox{\histw}{!}{\drawHistogram{0}{0}{4}{0}{1}{1}}
& {\footnotesize\num{ 0.0}\%} & \resizebox{\histw}{!}{\drawHistogram{0}{0}{5}{0}{1}{0}}
& {\footnotesize\num{ 0.0}\%} & \resizebox{\histw}{!}{\drawHistogram{0}{0}{4}{0}{2}{0}}
& {\footnotesize\num{ 0.0}\%} & \resizebox{\histw}{!}{\drawHistogram{0}{0}{5}{0}{1}{0}}
& {\footnotesize\num{ 16.666666666666664}\%} & \resizebox{\histw}{!}{\drawHistogram{0}{0}{2}{0}{3}{1}}
& {\footnotesize\num{ 13.88888888888889}\%} & \resizebox{\histw}{!}{\drawHistogram{0}{0}{23}{0}{8}{5}}
\\
{\small\texttt{qwen2.5:0.5b}}
& {\footnotesize\num{ 0.0}\%} & \resizebox{\histw}{!}{\drawHistogram{0}{0}{3}{0}{3}{0}}
& {\footnotesize\num{ 0.0}\%} & \resizebox{\histw}{!}{\drawHistogram{0}{0}{3}{0}{3}{0}}
& {\footnotesize\num{ 0.0}\%} & \resizebox{\histw}{!}{\drawHistogram{0}{1}{3}{0}{2}{0}}
& {\footnotesize\num{ 0.0}\%} & \resizebox{\histw}{!}{\drawHistogram{0}{0}{4}{0}{2}{0}}
& {\footnotesize\num{ 0.0}\%} & \resizebox{\histw}{!}{\drawHistogram{0}{0}{5}{0}{1}{0}}
& {\footnotesize\num{ 0.0}\%} & \resizebox{\histw}{!}{\drawHistogram{0}{1}{3}{0}{2}{0}}
& {\footnotesize\num{ 0.0}\%} & \resizebox{\histw}{!}{\drawHistogram{0}{2}{21}{0}{13}{0}}
\\
{\small\texttt{qwq:32b}}
& {\footnotesize\num{ 0.0}\%} & \resizebox{\histw}{!}{\drawHistogram{3}{3}{0}{0}{0}{0}}
& {\footnotesize\num{ 0.0}\%} & \resizebox{\histw}{!}{\drawHistogram{3}{3}{0}{0}{0}{0}}
& {\footnotesize\num{ 33.33333333333333}\%} & \resizebox{\histw}{!}{\drawHistogram{1}{3}{0}{0}{0}{2}}
& {\footnotesize\num{ 0.0}\%} & \resizebox{\histw}{!}{\drawHistogram{3}{3}{0}{0}{0}{0}}
& {\footnotesize\num{ 0.0}\%} & \resizebox{\histw}{!}{\drawHistogram{2}{4}{0}{0}{0}{0}}
& {\footnotesize\num{ 0.0}\%} & \resizebox{\histw}{!}{\drawHistogram{3}{3}{0}{0}{0}{0}}
& {\footnotesize\num{ 5.555555555555555}\%} & \resizebox{\histw}{!}{\drawHistogram{15}{19}{0}{0}{0}{2}}
\\

    \bottomrule
    \end{tabular}
    }
\end{table}

} 

\clearpage

\afterpage{

\clearpage

\begin{figure}[t!]
    \centering
    \resizebox{\textwidth}{!}{  
    \includegraphics[width=1\linewidth]{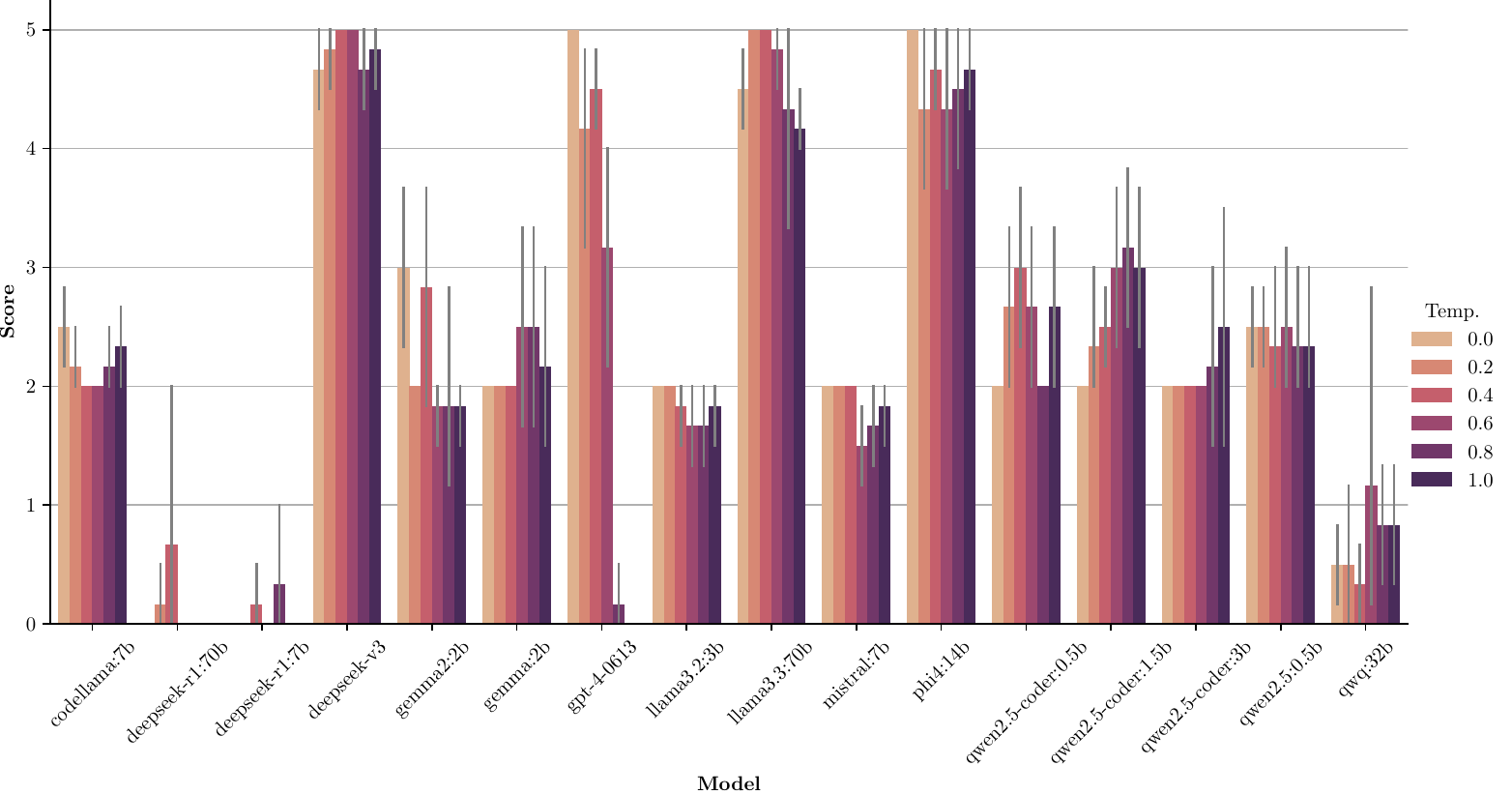}
    }
    \caption{\figurecaption{3}}
	\label{fig:res_prompt3}
\end{figure}

\begin{table}[h!]
    \caption{\tablecaption{3}} 
    \label{tab:res_prompt3}
    \resizebox{\textwidth}{!}{  
    \begin{tabular}{lr@{}rr@{}rr@{}rr@{}rr@{}rr@{}rr@{}r}
        \toprule
        \multirow{2.2}{*}{Model} & \multicolumn{14}{l}{Temperature} \\
              \cmidrule(l){2-15}
              & \multicolumn{2}{r}{0.0} & \multicolumn{2}{r}{0.2} & \multicolumn{2}{r}{0.4} & \multicolumn{2}{r}{0.6} & \multicolumn{2}{r}{0.8} & \multicolumn{2}{r}{1.0} & \multicolumn{2}{r}{Overall} \\
        \midrule

{\small\texttt{codellama:7b}}
& {\footnotesize\num{ 0.0}\%} & \resizebox{\histw}{!}{\drawHistogram{0}{0}{3}{3}{0}{0}}
& {\footnotesize\num{ 0.0}\%} & \resizebox{\histw}{!}{\drawHistogram{0}{0}{5}{1}{0}{0}}
& {\footnotesize\num{ 0.0}\%} & \resizebox{\histw}{!}{\drawHistogram{0}{0}{6}{0}{0}{0}}
& {\footnotesize\num{ 0.0}\%} & \resizebox{\histw}{!}{\drawHistogram{0}{0}{6}{0}{0}{0}}
& {\footnotesize\num{ 0.0}\%} & \resizebox{\histw}{!}{\drawHistogram{0}{0}{5}{1}{0}{0}}
& {\footnotesize\num{ 0.0}\%} & \resizebox{\histw}{!}{\drawHistogram{0}{0}{4}{2}{0}{0}}
& {\footnotesize\num{ 0.0}\%} & \resizebox{\histw}{!}{\drawHistogram{0}{0}{29}{7}{0}{0}}
\\
{\small\texttt{deepseek-r1:70b}}
& {\footnotesize\num{ 0.0}\%} & \resizebox{\histw}{!}{\drawHistogram{6}{0}{0}{0}{0}{0}}
& {\footnotesize\num{ 0.0}\%} & \resizebox{\histw}{!}{\drawHistogram{5}{1}{0}{0}{0}{0}}
& {\footnotesize\num{ 0.0}\%} & \resizebox{\histw}{!}{\drawHistogram{5}{0}{0}{0}{1}{0}}
& {\footnotesize\num{ 0.0}\%} & \resizebox{\histw}{!}{\drawHistogram{6}{0}{0}{0}{0}{0}}
& {\footnotesize\num{ 0.0}\%} & \resizebox{\histw}{!}{\drawHistogram{6}{0}{0}{0}{0}{0}}
& {\footnotesize\num{ 0.0}\%} & \resizebox{\histw}{!}{\drawHistogram{6}{0}{0}{0}{0}{0}}
& {\footnotesize\num{ 0.0}\%} & \resizebox{\histw}{!}{\drawHistogram{34}{1}{0}{0}{1}{0}}
\\
{\small\texttt{deepseek-r1:7b}}
& {\footnotesize\num{ 0.0}\%} & \resizebox{\histw}{!}{\drawHistogram{6}{0}{0}{0}{0}{0}}
& {\footnotesize\num{ 0.0}\%} & \resizebox{\histw}{!}{\drawHistogram{6}{0}{0}{0}{0}{0}}
& {\footnotesize\num{ 0.0}\%} & \resizebox{\histw}{!}{\drawHistogram{5}{1}{0}{0}{0}{0}}
& {\footnotesize\num{ 0.0}\%} & \resizebox{\histw}{!}{\drawHistogram{6}{0}{0}{0}{0}{0}}
& {\footnotesize\num{ 0.0}\%} & \resizebox{\histw}{!}{\drawHistogram{5}{0}{1}{0}{0}{0}}
& {\footnotesize\num{ 0.0}\%} & \resizebox{\histw}{!}{\drawHistogram{6}{0}{0}{0}{0}{0}}
& {\footnotesize\num{ 0.0}\%} & \resizebox{\histw}{!}{\drawHistogram{34}{1}{1}{0}{0}{0}}
\\
{\small\texttt{deepseek-v3}}
& {\footnotesize\num{ 66.66666666666666}\%} & \resizebox{\histw}{!}{\drawHistogram{0}{0}{0}{0}{2}{4}}
& {\footnotesize\num{ 83.33333333333334}\%} & \resizebox{\histw}{!}{\drawHistogram{0}{0}{0}{0}{1}{5}}
& {\footnotesize\num{ 100.0}\%} & \resizebox{\histw}{!}{\drawHistogram{0}{0}{0}{0}{0}{6}}
& {\footnotesize\num{ 100.0}\%} & \resizebox{\histw}{!}{\drawHistogram{0}{0}{0}{0}{0}{6}}
& {\footnotesize\num{ 66.66666666666666}\%} & \resizebox{\histw}{!}{\drawHistogram{0}{0}{0}{0}{2}{4}}
& {\footnotesize\num{ 83.33333333333334}\%} & \resizebox{\histw}{!}{\drawHistogram{0}{0}{0}{0}{1}{5}}
& {\footnotesize\num{ 83.33333333333334}\%} & \resizebox{\histw}{!}{\drawHistogram{0}{0}{0}{0}{6}{30}}
\\
{\small\texttt{gemma2:2b}}
& {\footnotesize\num{ 0.0}\%} & \resizebox{\histw}{!}{\drawHistogram{0}{0}{3}{0}{3}{0}}
& {\footnotesize\num{ 0.0}\%} & \resizebox{\histw}{!}{\drawHistogram{0}{0}{6}{0}{0}{0}}
& {\footnotesize\num{ 0.0}\%} & \resizebox{\histw}{!}{\drawHistogram{0}{1}{2}{0}{3}{0}}
& {\footnotesize\num{ 0.0}\%} & \resizebox{\histw}{!}{\drawHistogram{0}{1}{5}{0}{0}{0}}
& {\footnotesize\num{ 0.0}\%} & \resizebox{\histw}{!}{\drawHistogram{0}{3}{2}{0}{1}{0}}
& {\footnotesize\num{ 0.0}\%} & \resizebox{\histw}{!}{\drawHistogram{0}{1}{5}{0}{0}{0}}
& {\footnotesize\num{ 0.0}\%} & \resizebox{\histw}{!}{\drawHistogram{0}{6}{23}{0}{7}{0}}
\\
{\small\texttt{gemma:2b}}
& {\footnotesize\num{ 0.0}\%} & \resizebox{\histw}{!}{\drawHistogram{0}{0}{6}{0}{0}{0}}
& {\footnotesize\num{ 0.0}\%} & \resizebox{\histw}{!}{\drawHistogram{0}{0}{6}{0}{0}{0}}
& {\footnotesize\num{ 0.0}\%} & \resizebox{\histw}{!}{\drawHistogram{0}{0}{6}{0}{0}{0}}
& {\footnotesize\num{ 0.0}\%} & \resizebox{\histw}{!}{\drawHistogram{0}{1}{3}{0}{2}{0}}
& {\footnotesize\num{ 0.0}\%} & \resizebox{\histw}{!}{\drawHistogram{0}{1}{3}{0}{2}{0}}
& {\footnotesize\num{ 0.0}\%} & \resizebox{\histw}{!}{\drawHistogram{0}{1}{4}{0}{1}{0}}
& {\footnotesize\num{ 0.0}\%} & \resizebox{\histw}{!}{\drawHistogram{0}{3}{28}{0}{5}{0}}
\\
{\small\texttt{gpt-4-0613}}
& {\footnotesize\num{ 100.0}\%} & \resizebox{\histw}{!}{\drawHistogram{0}{0}{0}{0}{0}{6}}
& {\footnotesize\num{ 50.0}\%} & \resizebox{\histw}{!}{\drawHistogram{0}{0}{1}{0}{2}{3}}
& {\footnotesize\num{ 50.0}\%} & \resizebox{\histw}{!}{\drawHistogram{0}{0}{0}{0}{3}{3}}
& {\footnotesize\num{ 0.0}\%} & \resizebox{\histw}{!}{\drawHistogram{0}{1}{1}{0}{4}{0}}
& {\footnotesize\num{ 0.0}\%} & \resizebox{\histw}{!}{\drawHistogram{5}{1}{0}{0}{0}{0}}
& {\footnotesize\num{ 0.0}\%} & \resizebox{\histw}{!}{\drawHistogram{6}{0}{0}{0}{0}{0}}
& {\footnotesize\num{ 33.33333333333333}\%} & \resizebox{\histw}{!}{\drawHistogram{11}{2}{2}{0}{9}{12}}
\\
{\small\texttt{llama3.2:3b}}
& {\footnotesize\num{ 0.0}\%} & \resizebox{\histw}{!}{\drawHistogram{0}{0}{6}{0}{0}{0}}
& {\footnotesize\num{ 0.0}\%} & \resizebox{\histw}{!}{\drawHistogram{0}{0}{6}{0}{0}{0}}
& {\footnotesize\num{ 0.0}\%} & \resizebox{\histw}{!}{\drawHistogram{0}{1}{5}{0}{0}{0}}
& {\footnotesize\num{ 0.0}\%} & \resizebox{\histw}{!}{\drawHistogram{0}{2}{4}{0}{0}{0}}
& {\footnotesize\num{ 0.0}\%} & \resizebox{\histw}{!}{\drawHistogram{0}{2}{4}{0}{0}{0}}
& {\footnotesize\num{ 0.0}\%} & \resizebox{\histw}{!}{\drawHistogram{0}{1}{5}{0}{0}{0}}
& {\footnotesize\num{ 0.0}\%} & \resizebox{\histw}{!}{\drawHistogram{0}{6}{30}{0}{0}{0}}
\\
{\small\texttt{llama3.3:70b}}
& {\footnotesize\num{ 50.0}\%} & \resizebox{\histw}{!}{\drawHistogram{0}{0}{0}{0}{3}{3}}
& {\footnotesize\num{ 100.0}\%} & \resizebox{\histw}{!}{\drawHistogram{0}{0}{0}{0}{0}{6}}
& {\footnotesize\num{ 100.0}\%} & \resizebox{\histw}{!}{\drawHistogram{0}{0}{0}{0}{0}{6}}
& {\footnotesize\num{ 83.33333333333334}\%} & \resizebox{\histw}{!}{\drawHistogram{0}{0}{0}{0}{1}{5}}
& {\footnotesize\num{ 66.66666666666666}\%} & \resizebox{\histw}{!}{\drawHistogram{0}{0}{1}{0}{1}{4}}
& {\footnotesize\num{ 16.666666666666664}\%} & \resizebox{\histw}{!}{\drawHistogram{0}{0}{0}{0}{5}{1}}
& {\footnotesize\num{ 69.44444444444444}\%} & \resizebox{\histw}{!}{\drawHistogram{0}{0}{1}{0}{10}{25}}
\\
{\small\texttt{mistral:7b}}
& {\footnotesize\num{ 0.0}\%} & \resizebox{\histw}{!}{\drawHistogram{0}{0}{6}{0}{0}{0}}
& {\footnotesize\num{ 0.0}\%} & \resizebox{\histw}{!}{\drawHistogram{0}{0}{6}{0}{0}{0}}
& {\footnotesize\num{ 0.0}\%} & \resizebox{\histw}{!}{\drawHistogram{0}{0}{6}{0}{0}{0}}
& {\footnotesize\num{ 0.0}\%} & \resizebox{\histw}{!}{\drawHistogram{0}{3}{3}{0}{0}{0}}
& {\footnotesize\num{ 0.0}\%} & \resizebox{\histw}{!}{\drawHistogram{0}{2}{4}{0}{0}{0}}
& {\footnotesize\num{ 0.0}\%} & \resizebox{\histw}{!}{\drawHistogram{0}{1}{5}{0}{0}{0}}
& {\footnotesize\num{ 0.0}\%} & \resizebox{\histw}{!}{\drawHistogram{0}{6}{30}{0}{0}{0}}
\\
{\small\texttt{phi4:14b}}
& {\footnotesize\num{ 100.0}\%} & \resizebox{\histw}{!}{\drawHistogram{0}{0}{0}{0}{0}{6}}
& {\footnotesize\num{ 66.66666666666666}\%} & \resizebox{\histw}{!}{\drawHistogram{0}{0}{0}{2}{0}{4}}
& {\footnotesize\num{ 66.66666666666666}\%} & \resizebox{\histw}{!}{\drawHistogram{0}{0}{0}{0}{2}{4}}
& {\footnotesize\num{ 66.66666666666666}\%} & \resizebox{\histw}{!}{\drawHistogram{0}{0}{0}{2}{0}{4}}
& {\footnotesize\num{ 66.66666666666666}\%} & \resizebox{\histw}{!}{\drawHistogram{0}{0}{0}{1}{1}{4}}
& {\footnotesize\num{ 66.66666666666666}\%} & \resizebox{\histw}{!}{\drawHistogram{0}{0}{0}{0}{2}{4}}
& {\footnotesize\num{ 72.22222222222221}\%} & \resizebox{\histw}{!}{\drawHistogram{0}{0}{0}{5}{5}{26}}
\\
{\small\texttt{qwen2.5-coder:0.5b}}
& {\footnotesize\num{ 0.0}\%} & \resizebox{\histw}{!}{\drawHistogram{0}{0}{6}{0}{0}{0}}
& {\footnotesize\num{ 0.0}\%} & \resizebox{\histw}{!}{\drawHistogram{0}{0}{4}{0}{2}{0}}
& {\footnotesize\num{ 0.0}\%} & \resizebox{\histw}{!}{\drawHistogram{0}{0}{3}{0}{3}{0}}
& {\footnotesize\num{ 0.0}\%} & \resizebox{\histw}{!}{\drawHistogram{0}{0}{4}{0}{2}{0}}
& {\footnotesize\num{ 0.0}\%} & \resizebox{\histw}{!}{\drawHistogram{0}{0}{6}{0}{0}{0}}
& {\footnotesize\num{ 0.0}\%} & \resizebox{\histw}{!}{\drawHistogram{0}{0}{4}{0}{2}{0}}
& {\footnotesize\num{ 0.0}\%} & \resizebox{\histw}{!}{\drawHistogram{0}{0}{27}{0}{9}{0}}
\\
{\small\texttt{qwen2.5-coder:1.5b}}
& {\footnotesize\num{ 0.0}\%} & \resizebox{\histw}{!}{\drawHistogram{0}{0}{6}{0}{0}{0}}
& {\footnotesize\num{ 0.0}\%} & \resizebox{\histw}{!}{\drawHistogram{0}{0}{5}{0}{1}{0}}
& {\footnotesize\num{ 0.0}\%} & \resizebox{\histw}{!}{\drawHistogram{0}{0}{3}{3}{0}{0}}
& {\footnotesize\num{ 0.0}\%} & \resizebox{\histw}{!}{\drawHistogram{0}{0}{2}{2}{2}{0}}
& {\footnotesize\num{ 0.0}\%} & \resizebox{\histw}{!}{\drawHistogram{0}{0}{2}{1}{3}{0}}
& {\footnotesize\num{ 0.0}\%} & \resizebox{\histw}{!}{\drawHistogram{0}{0}{2}{2}{2}{0}}
& {\footnotesize\num{ 0.0}\%} & \resizebox{\histw}{!}{\drawHistogram{0}{0}{20}{8}{8}{0}}
\\
{\small\texttt{qwen2.5-coder:3b}}
& {\footnotesize\num{ 0.0}\%} & \resizebox{\histw}{!}{\drawHistogram{0}{0}{6}{0}{0}{0}}
& {\footnotesize\num{ 0.0}\%} & \resizebox{\histw}{!}{\drawHistogram{0}{0}{6}{0}{0}{0}}
& {\footnotesize\num{ 0.0}\%} & \resizebox{\histw}{!}{\drawHistogram{0}{0}{6}{0}{0}{0}}
& {\footnotesize\num{ 0.0}\%} & \resizebox{\histw}{!}{\drawHistogram{0}{0}{6}{0}{0}{0}}
& {\footnotesize\num{ 0.0}\%} & \resizebox{\histw}{!}{\drawHistogram{0}{1}{4}{0}{1}{0}}
& {\footnotesize\num{ 0.0}\%} & \resizebox{\histw}{!}{\drawHistogram{0}{2}{1}{1}{2}{0}}
& {\footnotesize\num{ 0.0}\%} & \resizebox{\histw}{!}{\drawHistogram{0}{3}{29}{1}{3}{0}}
\\
{\small\texttt{qwen2.5:0.5b}}
& {\footnotesize\num{ 0.0}\%} & \resizebox{\histw}{!}{\drawHistogram{0}{0}{3}{3}{0}{0}}
& {\footnotesize\num{ 0.0}\%} & \resizebox{\histw}{!}{\drawHistogram{0}{0}{3}{3}{0}{0}}
& {\footnotesize\num{ 0.0}\%} & \resizebox{\histw}{!}{\drawHistogram{0}{0}{5}{0}{1}{0}}
& {\footnotesize\num{ 0.0}\%} & \resizebox{\histw}{!}{\drawHistogram{0}{0}{4}{1}{1}{0}}
& {\footnotesize\num{ 0.0}\%} & \resizebox{\histw}{!}{\drawHistogram{0}{0}{5}{0}{1}{0}}
& {\footnotesize\num{ 0.0}\%} & \resizebox{\histw}{!}{\drawHistogram{0}{0}{5}{0}{1}{0}}
& {\footnotesize\num{ 0.0}\%} & \resizebox{\histw}{!}{\drawHistogram{0}{0}{25}{7}{4}{0}}
\\
{\small\texttt{qwq:32b}}
& {\footnotesize\num{ 0.0}\%} & \resizebox{\histw}{!}{\drawHistogram{3}{3}{0}{0}{0}{0}}
& {\footnotesize\num{ 0.0}\%} & \resizebox{\histw}{!}{\drawHistogram{4}{1}{1}{0}{0}{0}}
& {\footnotesize\num{ 0.0}\%} & \resizebox{\histw}{!}{\drawHistogram{4}{2}{0}{0}{0}{0}}
& {\footnotesize\num{ 16.666666666666664}\%} & \resizebox{\histw}{!}{\drawHistogram{3}{2}{0}{0}{0}{1}}
& {\footnotesize\num{ 0.0}\%} & \resizebox{\histw}{!}{\drawHistogram{2}{3}{1}{0}{0}{0}}
& {\footnotesize\num{ 0.0}\%} & \resizebox{\histw}{!}{\drawHistogram{2}{3}{1}{0}{0}{0}}
& {\footnotesize\num{ 2.7777777777777777}\%} & \resizebox{\histw}{!}{\drawHistogram{18}{14}{3}{0}{0}{1}}
\\

        \bottomrule
    \end{tabular}
    }
\end{table}

} 

\clearpage
\clearpage

Prompt~3, while similar to Prompt~2 in many respects, requires the requested function to return a concrete power loss value rather than merely the index of the position with the lowest loss. This distinction arguably makes it the most complex task for the models evaluated in this study. The~results for this prompt are presented in Figure~\ref{fig:res_prompt3} and Table~\ref{tab:res_prompt3}. The~same four models continued to generate accurate code consistently, though~within a more limited range of temperature settings. DeepSeek-V3 demonstrated the highest overall consistency, reliably producing correct code at temperatures of $0.8$ ($2 \times 0.4$) and $1.2$ ($2 \times 0.6)$, while maintaining a high percentage of accurate responses across the remaining temperatures. GPT-4 and Phi-4 achieved 100\% accuracy when the temperature was set to zero. However, while Phi-4 remained highly consistent at higher temperatures, GPT-4 exhibited a significant decline in performance. LLaMA-3.3 also demonstrated strong consistency, achieving 100\% accuracy in all runs at temperatures of 0.2 and 0.4. None of the remaining models were able to successfully complete this task. The~only exception was Qwen's QwQ (32B), which generated a single correct response at a temperature of 0.6. However, beyond~this isolated instance, it predominantly produced code containing invalid syntax or runtime~errors.

\begin{figure}[H]
    \centering
    \includegraphics[width=0.85\linewidth]{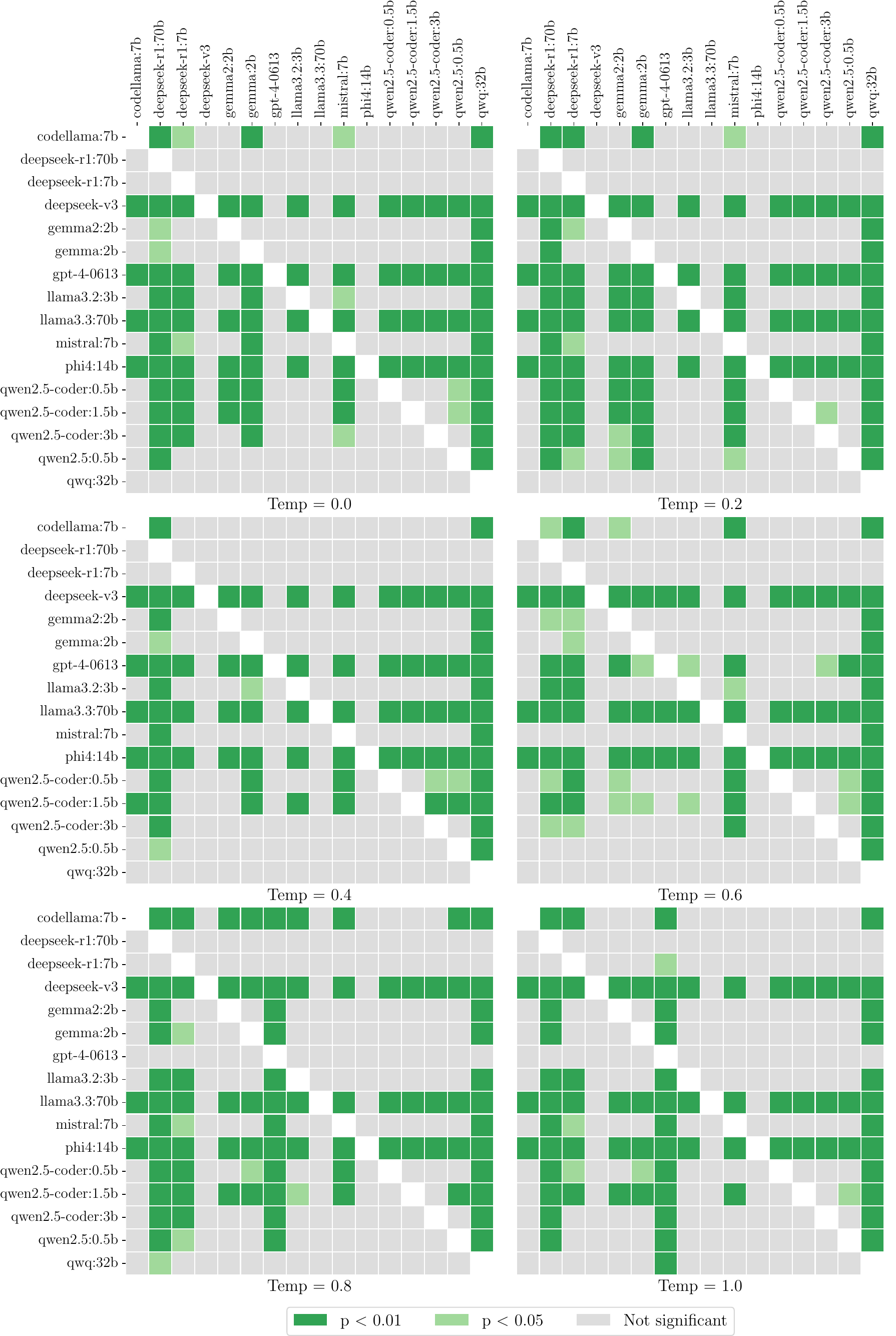}
    \caption{Pairwise significance heatmap of model performance comparisons for the three prompts across temperatures. Each colored block represents the $p$-value of a one-sided, rank-based stratified permutation test between two models (model in row vs. model in column) for a given temperature. Cells are colored based on statistical significance after Benjamini--Hochberg FDR multiple testing correction: dark green indicates a significant advantage of the model in the row against the model in the column ($p < 0.01$), light green indicates moderate significant advantage ($p < 0.05$), and~light gray denotes no significant difference. Temperatures for online models, \texttt{deepseek-v3} and \texttt{gpt-4-0613}, are twice the displayed~values.}
	\label{fig:pvalues}
\end{figure}

Figure~\ref{fig:pvalues} presents a pairwise significance heatmap based on $p$-values from a stratified permutation test, after~FDR multiple testing correction, indicating which models (in rows) statistically outperformed others (in columns) across temperatures. Table~\ref{tab:significant_wins} summarizes these results, showing the number of models each system significantly outperformed at each temperature, as~well as the overall total across all temperatures. These results reinforce what was observed in the descriptive statistics---namely that DeepSeek-V3, GPT-4, Phi-4, and~LLaMA-3.3 are the most consistent and competitive models in these engineering tasks. At~nearly all temperature levels, these models significantly outperformed the majority of alternatives, with~corrected $p$-values below the 0.05 threshold in a substantial number of pairwise comparisons. In~particular, DeepSeek-V3, Phi-4, and~LLaMA-3.3 achieved the highest number of significant wins at every temperature, while GPT-4 showed similarly strong performance at lower temperatures but exhibited a sharp decline in statistical superiority as temperature~increased.

In contrast, the~two DeepSeek-R1 models, as~well as QwQ, registered very few significant wins at any temperature. Crucially, their only advantages were against GPT-4 at higher temperatures, where its output becomes increasingly random and unsuitable for these types of coding tasks. This further confirms their limited effectiveness, as~already observed in previous results. An~additional insight---less apparent in the descriptive statistics but clearly highlighted in Table~\ref{tab:significant_wins}---is the lack of correlation between model size and performance within the Qwen coder family. Specifically, the~1.5B Qwen coder model achieved the fourth highest total number of pairwise wins (47), surpassing even GPT-4 (45), while the larger 3B variant achieved roughly half as~many.

Finally, Figure~\ref{fig:scores_by_prompt} presents the mean scores for the tested models across all three prompts, aggregating results from all seeds and temperature settings. While the initial assumption was that Prompts 1 to 3 increase in complexity, and~the results thus far appear to support this hypothesis, Figure~\ref{fig:scores_by_prompt} provides a more comprehensive perspective. For~most models, the~mean score declines progressively with increasing prompt complexity, reinforcing this assumption. However, exceptions include both Gemma models and the non-coder Qwen-2.5 model, where the score reduction is not strictly monotonic. Another observation from this figure is that the highest performing models---DeepSeek-V3, GPT-4, LLaMA-3.3, and~Phi-4---maintain consistent performance across prompts, with~only a slight decline in mean score as complexity~increases.

\afterpage{
\clearpage

\begin{table}[htbp!]
\footnotesize
\caption{Number of statistically significant pairwise wins (corrected $p < 0.05$) per model across temperature settings. Bold values indicate the highest number of wins for each temperature column (including ties). Each cell represents how many times a given model significantly outperformed others at the corresponding temperature. Temperatures for online models, \texttt{deepseek-v3} and \texttt{gpt-4-0613}, are twice the displayed values.}
\label{tab:significant_wins}
\begin{tabular}{lrrrrrrr}
    \toprule
    \multirow{2.2}{*}{Model} & \multicolumn{6}{l}{Temperature}\\
    \cmidrule(l){2-8}
    & 0.0 & 0.2 & 0.4 & 0.6 & 0.8 & 1.0 & Overall \\
    \midrule
    \texttt{codellama:7b} & 5 & 5 & 2 & 5 & 9 & 4 & 30 \\
    \texttt{deepseek-r1:70b} & 0 & 0 & 0 & 0 & 0 & 0 & 0 \\
    \texttt{deepseek-r1:7b} & 0 & 0 & 0 & 0 & 0 & 1 & 1 \\
    \texttt{deepseek-v3} & \textbf{12} & \textbf{12} & \textbf{12} & \textbf{13} & \textbf{13} & \textbf{13} & \textbf{75} \\
    \texttt{gemma2:2b} & 2 & 3 & 2 & 3 & 3 & 3 & 16 \\
    \texttt{gemma:2b} & 2 & 2 & 2 & 2 & 4 & 3 & 15 \\
    \texttt{gpt-4-0613} & \textbf{12} & \textbf{12} & \textbf{12} & 9 & 0 & 0 & 45\\
    \texttt{llama3.2:3b} & 5 & 6 & 3 & 4 & 4 & 4 & 26\\
    \texttt{llama3.3:70b} & \textbf{12} & \textbf{12} & \textbf{12} & \textbf{13} & \textbf{13} & \textbf{13} & \textbf{75}\\
    \texttt{mistral:7b} & 4 & 3 & 2 & 1 & 4 & 4 & 18\\
    \texttt{phi4:14b} & \textbf{12} & \textbf{12} & \textbf{12} & \textbf{13} & \textbf{13} & \textbf{13} & \textbf{75} \\
    \texttt{qwen2.5-coder:0.5b} & 7 & 6 & 6 & 6 & 6 & 5 & 36 \\
    \texttt{qwen2.5-coder:1.5b} & 7 & 7 & 8 & 8 & 9 & 8 & 47 \\
    \texttt{qwen2.5-coder:3b} & 5 & 6 & 2 & 4 & 4 & 3 & 24 \\
    \texttt{qwen2.5:0.5b} & 2 & 6 & 2 & 1 & 4 & 3 & 18\\
    \texttt{qwq:32b} & 0 & 0 & 0 & 0 & 1 & 1 & 2\\
    \bottomrule
\end{tabular}

\end{table}

\begin{figure}[htbp!]
    \centering
    \includegraphics[width=1\linewidth]{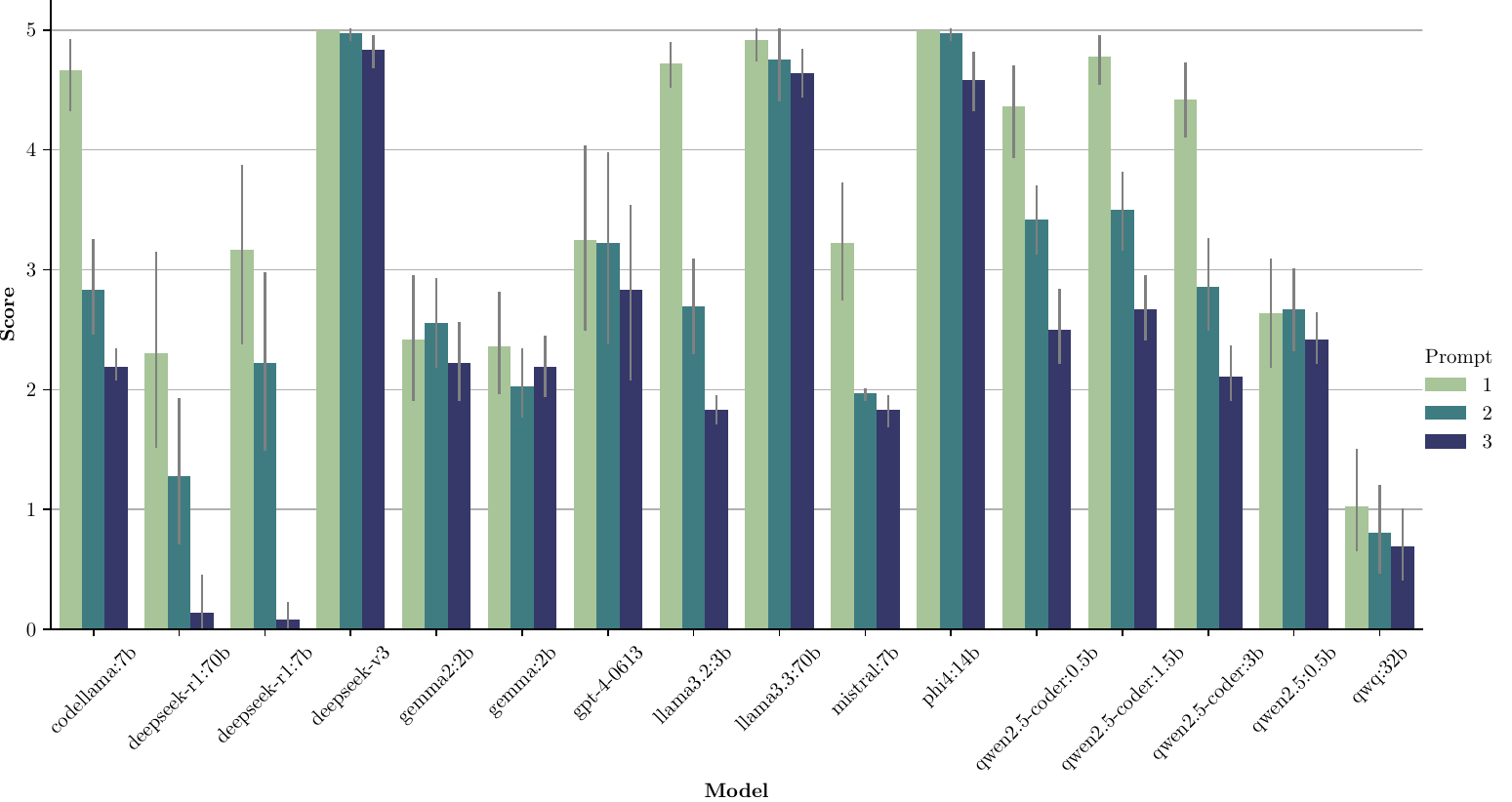}
    \caption{Mean answer score for the tested models and the three prompts. Each combination of model and prompt was tested 36 times (6 seeds $\times$ 6 temperatures). Error bars denote a 95\% confidence interval.}
	\label{fig:scores_by_prompt}
\end{figure}
} 

\clearpage
\clearpage

\section{Discussion}
\label{sec:discussion}

Within the DeepSeek model family, there was a surprising discrepancy between the well-performing DeepSeek-V3 and the underperforming DeepSeek-R1 models. The DeepSeek-R1 versions (7B and 70B), despite their larger parameter counts, rarely generated correct code. Interestingly, the~DeepSeek-R1 models, as~well as Qwen's QwQ (32B), tended to generate answers over five times longer than those from other models, yet without improved correctness. While these verbose outputs are particularly noticeable, we did not investigate the reasons behind them because this lies beyond the scope of this study. Nonetheless, the~generated data---as well as further analyses on this matter---are available on Zenodo\footnote{\url{https://doi.org/10.5281/zenodo.14888673}} 
 and may be addressed in future studies.

An important observation is that GPT-4 exhibits essentially random outputs when operating at higher temperatures. This behavior aligns with OpenAI's own documentation, which indicates that temperatures above 1.2 or 1.4 may lead to increasingly stochastic completions. In~contrast, the~other top-performing models in this study---DeepSeek-V3, LLaMA-3.3, and~Phi-4---remain relatively robust under higher temperature settings. These considerations indicate that temperature influences each model differently. Differences in temperature scaling ranges (0--1 vs.\ 0--2) further complicate direct~comparisons.

Although one might expect a clear correlation between model size and code generation quality, results support a more involved situation among locally run models. Larger models such as DeepSeek-R1 (70B) and QwQ (32B) do not necessarily outperform smaller alternatives: their answers were typically long yet largely incorrect. Conversely, some mid- to large-scale models, such as Phi-4 (14B) and LLaMA-3.3 (70B), consistently provided accurate solutions to all prompts. Another example, LLaMA-3.2 (3B), showed reasonable performance for simpler tasks but struggled with more complex prompts, highlighting a lower boundary for parameter count beyond which performance degrades. In~contrast, Qwen's smaller coder models (0.5B, 1.5B, 3B) did not show any clear advantage with increasing size, confirming that raw parameter counts alone are insufficient to predict success across different~tasks.

Within the Gemini-based lineage, Gemma-2 offered marginal improvements over its older v1.1 sibling, though~neither model consistently produced correct outputs. On~the other hand, LLaMA-3.3 (70B) clearly outperformed the related LLaMA-3.2 (3B), a~result likely driven by its substantially larger parameter count. Phi-4 merits special mention for delivering accurate code across all tasks, seeds, and~temperatures, while requiring considerably fewer parameters (14B) than the largest competitors. This affords Phi-4 a strong performance/size ratio among the locally executed~models.

To support these observations, a~stratified permutation test with FDR correction was applied across all model pairs and temperatures. The~resulting significance heatmap and win counts showed strong agreement with the descriptive statistics. DeepSeek-V3, \mbox{Phi-4,} and~LLaMA-3.3 consistently achieved the highest number of statistically significant wins, while GPT-4 also dominated at lower temperatures. These results reinforce that the observed differences in model performance are statistically meaningful and not artifacts of randomness or scoring~variability.

From a broader perspective, these findings support the notion that carefully tuned, locally run models can achieve near-state-of-the-art performance in specialized Python code generation tasks without necessarily relying on proprietary solutions. Specifically, both Phi-4 and LLaMA-3.3 proved capable of reliably generating correct solutions for the type of UAV/LoRaWAN planning prompts tested in this work. Their consistency in providing accurate answers under varying seeds and temperature conditions places them among the top-performing models overall, comparable to GPT-4 and DeepSeek-V3. These results address the central research question: lightweight and locally executed LLMs can, in~fact, generate correct Python code for relatively simple LoRaWAN and UAV planning tasks, provided that their parameter counts and training procedures meet a certain threshold of quality and scale. The~performance of Phi-4 was particularly impressive, especially considering it is a relatively lightweight~model.

\section{Limitations}
\label{sec:limitations}

Despite the insights gained from this study, several limitations should be acknowledged. First, the~selection of models, while diverse, was not exhaustive. Only a subset of locally run lightweight models was evaluated, and~online testing was limited to GPT-4 and DeepSeek-V3. Several potentially relevant models, such as Claude, Mistral (larger online versions), and~specialized coding models (e.g., Gemma Coder or DeepSeek Coder), were not included. This restricted scope leaves open the possibility that other models may perform competitively or even outperform those tested in this~study.

Second, model outputs were assessed solely based on functional correctness, without~a detailed qualitative analysis of the responses. This introduces the risk that some answers classified as correct may not have been genuinely derived but instead relied on unintended memorization, dataset leakage, or~other forms of `cheating'. While this concern is most relevant for Prompts 1 and 2, where only an index is returned, Prompt 3 mitigates this issue by requiring a real-valued output. Nevertheless, a~more rigorous analysis of response quality---including potential hallucinations, redundant reasoning, and~incorrect assumptions---would strengthen future~work.

Third, the~study relied on a single test case per function, which limits the robustness of correctness assessments. A~more comprehensive evaluation would include multiple test cases per function, ensuring that responses generalize beyond a specific input scenario. This is particularly relevant given the stochastic nature of LLM-generated code, where seemingly minor variations in the prompt or execution conditions can lead to significant changes in output~validity.

Fourth, all evaluations were conducted using zero-shot natural language prompts, without~fine-tuning or explicit prompt engineering. While this choice aligns with practical use cases where domain experts may rely on straightforward instructions, further experimentation with prompt optimization strategies---such as chain-of-thought prompting or few-shot learning---could provide deeper insights into model~capabilities.

Additionally, the~study focused on relatively simple UAV/LoRaWAN planning tasks. While these scenarios are relevant to real-world applications, they do not necessarily capture the full complexity of autonomous UAV coordination, network interference, or~real-time decision-making in dynamic environments. The~strong performance of top models suggests they may be capable of handling more complex scenarios, but~this remains an open question for future~research.

A final limitation concerns the use of statistical significance testing. While stratified permutation tests confirmed the robustness of performance differences, they do not account for the magnitude or practical implications of those differences. Moreover, the~use of discrete, ordinal scores simplifies model outputs and may obscure subtle qualitative distinctions. Although~multiple testing correction was applied to reduce false positives, this also reduces sensitivity to borderline effects. Additionally, comparisons at non-zero temperatures should be interpreted with caution, as~temperature scaling is handled differently across models, potentially resulting in varying degrees of output randomness for the same nominal value. These tests therefore complement, but~do not replace, the~broader descriptive analysis presented~earlier.

\clearpage
These limitations do not diminish the validity of the study's conclusions but highlight areas for refinement in subsequent investigations. A~broader model selection, more rigorous evaluation metrics, and~extended task complexity would further improve the understanding of LLMs' capabilities in UAV and LoRaWAN-related computational~tasks.

\section{Conclusions}
\label{sec:conclusions}

This paper analyzed the capabilities of 16 LLMs to generate Python functions for practical LoRaWAN-related engineering tasks involving UAV placement and signal propagation. By~progressively increasing the complexity of prompts, we evaluated each model's ability to return valid and correct solutions under a standardized scoring system. The~findings indicate that several recent models---particularly DeepSeek-V3, GPT-4, LLaMA-3.3, and~Phi-4---consistently generated accurate and executable functions. Particularly, Phi-4 displayed exceptional performance despite its relatively lightweight architecture, demonstrating that well-optimized, smaller-scale models can be highly effective for specialized engineering applications. Models that did not achieve high scores often struggled with prompt interpretation, code syntax, or~domain-specific computations, underlining the need for careful prompt engineering and model fine-tuning in similar~applications.

The demonstrated viability of lightweight and locally executed LLMs for specialized engineering tasks such as UAV planning in LoRaWAN environments suggests that these models could significantly lower computational barriers and costs, allowing for broader and more flexible integration of AI-driven code generation into practical engineering~workflows.

While this study highlighted the strong potential of LLMs in engineering workflows, certain limitations must be acknowledged, including the constrained model selection, the~single test case per function, and~the absence of qualitative analysis of responses. However, these limitations present opportunities for future research. Expanding test sets, incorporating more complex domain requirements, and~evaluating additional models---particularly other lightweight alternatives---could further enrich our understanding of LLM-driven code generation in wireless communications and related fields. Future research could also explore the incorporation of reinforcement learning with human feedback to further improve the code generation capabilities of lightweight LLMs~\cite{wong2024aligning}.

\section*{Data Availability}

The data generated by this study and its respective analysis are available at \url{https://doi.org/10.5281/zenodo.14888673} under the CC-BY license.

\section*{Acknowledgements}

This research was partially funded by: Funda\c c\~ao para a Ci\^encia e a Tecnologia (FCT, \url{https://ror.org/00snfqn58}) under Grants Copelabs ref. UIDB/04111/2020, Centro de Tecnologias e Sistemas (CTS) ref. UIDB/00066/2020, LASIGE Research Unit ref. UIDB/00408/2025, and COFAC ref. CEECINST/00002/2021/CP2788/CT0001; Instituto Lusófono de Investigação e Desenvolvimento (ILIND, Portugal) under~Project COFAC/ILIND\-/\-COPELABS\-/1/2024; and, Ministerio de Ciencia, Innovación y Universidades (MICIU/AEI/10.13039/501100011033, \url{https://ror.org/05r0vyz12}) under Project PID2023-147409NB-C21.
\bibliographystyle{elsarticle-num}

\end{document}